\documentclass[3p, preprint]{elsarticle}

\usepackage{graphicx}
\usepackage{amsmath}
\usepackage{amssymb}
\usepackage{mathtools}
\usepackage{epsfig,natbib,listings,lineno,xcolor,bm,algorithm,url,amsthm,psfrag}
\usepackage[normalem]{ulem}
\usepackage{subfig}
\usepackage{float}
\usepackage{algpseudocode}
\usepackage{subfig}
\usepackage{enumitem}
\usepackage{ulem}

\newtheorem{theorem}{Theorem}[section]

\newtheorem{lemma}[theorem]{Lemma}
\newtheorem{remark}{Remark}[section]

\DeclareMathOperator*{\argmin}{arg\,min}
\newcommand{\alec}[1]{\textcolor{black}{#1}}
\newcommand{\heather}[1]{\textcolor{black}{#1}}
\DeclarePairedDelimiter{\norm}{\lVert}{\rVert}

\title{Task-parallel \textit{in-situ} temporal compression of large-scale computational fluid dynamics data}

\author[1]{Heather Pacella\corref{cor1}}
\author[2]{Alec Dunton}
\author[3]{Alireza Doostan}
\author[1]{Gianluca Iaccarino}

\cortext[cor1]{Corresponding author, hpacella@stanford.edu}
\address[1]{Department of Mechanical Engineering, Stanford University, Stanford, CA 94305}
\address[2]{Department of Applied Mathematics, University of Colorado Boulder, Boulder, CO 80309}
\address[3]{Smead Aerospace Engineering Sciences, University of Colorado Boulder, Boulder, CO 80309}

\begin{document}

\begin{abstract}
   Present day computational fluid dynamics (CFD) simulations generate extremely large amounts of data, \alec{sometimes on the order of TB/s}. Often, a significant fraction of this data is discarded because current storage systems are unable to keep pace. To address this, data compression algorithms can be applied to data arrays containing flow quantities of interest (QoIs) to reduce the overall amount of storage. Compression methods either exactly reconstruct the original dataset (lossless compression) or provide an approximate representation of the original dataset (lossy compression). The matrix column \textit{interpolative decomposition} (ID) can be implemented as a type of lossy compression for data matrices that factors the original data matrix into a product of two smaller {\it factor matrices}. One of these matrices consists of a subset of the columns of the original data matrix, while the other is a coefficient matrix which approximates the columns of the original data matrix as linear combinations of the selected columns. Motivating this work is the observation that the structure of ID algorithms makes them a natural fit for the asynchronous nature of task-based parallelism; they are able to operate independently on sub-domains of the system of interest and, as a result, provide varied levels of compression. Using the task-based Legion programming model, a single-pass ID algorithm (SPID) for CFD applications is implemented. Performance studies, scalability, and the accuracy of the \alec{compression algorithms} are presented for a benchmark analytical Taylor-Green vortex problem, followed by a large-scale implementation of a compressible Taylor-Green vortex using a high-order Navier-Stokes solver. \alec{In both cases, compression factors exceeding 100 are achieved with relative errors at or below $10^{-3}$. Moreover, strong and weak scaling results demonstrate that introducing SPID to solvers leads to negligible increases in runtime.}
\end{abstract}


\maketitle

\section{Introduction and Motivation}
Advances in supercomputing over the past few decades have introduced a computational bottleneck. Soon-to-be-deployed Exascale computers are expected to offer a $1,000$-$10,000$-fold increase in floating point performance, but provide a relatively mere $10-100$-fold increase in available disk memory, working memory, and access speed~\cite{ang2012report}. This asymmetric technological progress gives rise to the following issue: floating point operations (FLOPs) are cheap, while memory, communication, and input/output (I/O) is not. This general trend has been emphasized in other reports such as~\cite{amarasinghe2009exascale,ashby2010opportunities,sprague2017turbulent,asch2018big,gerber2018crosscut} as well as in the design of \alec{next generation} Exascale systems~\cite{kunkel2014exascale}. 

The amount of data generated on these computers can easily exceed 1 TB/s \alec{in large-scale simulations of complex systems, e.g., flow control for wing design~\cite{rasquin2014scalable}}. If this is not reduced, storage systems will easily become overloaded and practitioners will not be able to use the \alec{simulation} for visualization, analysis, or other post-processing operations. This issue has inspired work in {\it data compression}, in which a memory-reduced version of an original dataset is stored, with the hope that the compressed format maintains an accurate representation of the original dataset.

\alec{Data} compression algorithms may be categorized as lossless or lossy. Lossless compression methods enable exact reconstruction of datasets, but with a small {\it compression factor}, defined as the ratio of the size of the original dataset to that of the compressed dataset. Due to the exorbitant size of data being generated on modern supercomputers, particularly for the category of physical simulations addressed in this work,  we do not explore lossless compression methods. For the interested reader, an extensive review of lossless compression techniques can be found in~\cite{li2018data}.

\alec{Lossy compression methods do not enable exact reconstruction of datasets.} They do, however, allow for much larger compression factors. Algorithms for lossy data compression rely on a diverse set of tools to construct memory-efficient representations of datasets. Such methods include bit truncation~\cite{gong2012multi} and predictive coding techniques. Notable examples of predictive coders include Compvox~\cite{fowler1994lossless}, SZ~\cite{di2016fast,tao2017significantly,liang2018error}, FPZIP (which supports both lossless and lossy compression~\cite{lindstrom2006fast}), Lorenzo~\cite{ibarria2003out}, and Isabela~\cite{lakshminarasimhan2011compressing,lehmann2014situ}. Transform-based methods entail the truncation of a set of coefficients obtained via, e.g., the discrete Legendre~\cite{otero2018lossy,marin2016large}, discrete cosine~\cite{yeo1995volume}, and wavelet transforms~\cite{cohen1992biorthogonal,farge1992wavelet,strang1996wavelets}. Other examples of transform-based algorithms for compression include ZFP~\cite{lindstrom2014fixed}, the Karhunen-Loeve transform~\cite{loeve1977elementary,therrien1992discrete}, and tensor approximations~\cite{hitchcock1927expression,tucker1966some,kroonenberg1980principal,de2000best,vannieuwenhoven2012new,austin2016parallel}. Recent work has also explored the potential for deep learning to enable {\it in situ} compression of turbulent flows~\cite{glaws2020deep}. In this present work, low-rank matrix approximations -- a transform based method -- are investigated for lossy compression.

\subsection{Contribution of this work}
Low-rank matrix methods have been used for the compression of large-scale simulation data in works such as~\cite{azaiez2019low}. Computing low-rank matrix approximations in which simulation data arrives into working memory online has been addressed \alec{in}~\cite{brand2006fast,zimmermann2018geometric,tropp2019streaming}. The present effort is focused on using low-rank matrix approximation to build {\it in situ} compression methods which exploit task-based parallelism to achieve high compression and concurrency on heterogeneous modern computing architectures.

To this end, we use the matrix interpolative decomposition (ID) for low-rank approximation~\cite{cheng2005compression,halko2011finding}. Our specific approaches follow directly from ID-based CFD compression algorithms presented in~\cite{dunton2020pass}. The matrix ID is inserted into and applied to data from computational fluid dynamics (CFD) applications using Legion~\cite{Legion_paper}. \alec{Legion is a data-centric parallel programming and runtime system designed for high performance computing applications. It allows for heterogeneous CFD-ID applications to be computationally efficient via asynchronous, task-based parallel execution.} We also detail how additional parallelism can be extracted though Legion's custom mapping interface, which allows the user to execute a single application code on multiple processor types. \alec{Although other algorithms, such as the randomized SVD~\cite{halko2011finding,yu2017single}, are also available for compression in simulation, we focus on deterministic ID methods for reasons discussed in Section~\ref{sec:ID}.}

The rest of the paper is organized as follows. In Section~\ref{sec:ID}, we provide background on and derivations of the matrix ID algorithms used for lossy data compression. In Section~\ref{sec:ID_task_based}, we outline how ID methods may be incorporated into a task-based parallel environment. In Section~\ref{sec:legion}, we outline the Legion programming model, which enables task-based parallelism. In Section~\ref{sec:numerical_experiments}, we present numerical experiments from applying the ID to (1) an analytical Taylor-Green vortex solution of the incompressible Navier-Stokes equations and (2) a large-scale implementation of a compressible Taylor-Green vortex using a high-order Navier-Stokes solver. In Section~\ref{sec:conclusions}, we draw conclusions and propose future avenues of research. Theoretical results relevant to Section~\ref{sec:ID} are provided in the Appendix.

\section{Interpolative decomposition}
\label{sec:ID}
\begin{figure}[!h]
       \center{\includegraphics[width=0.4\textwidth]{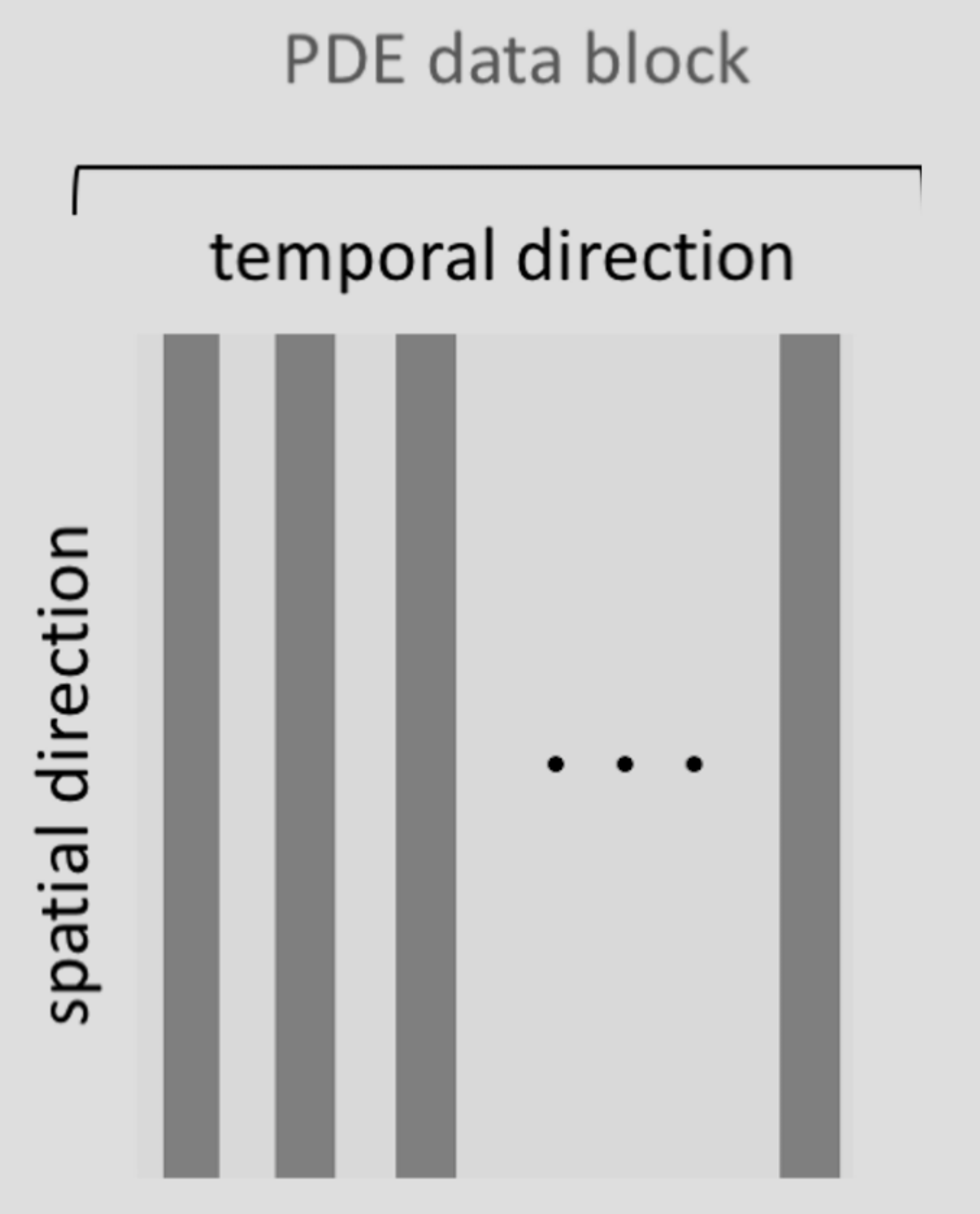}}
       \caption{\label{fig:PDE_data} Schematic of PDE data block. Columns constitute vectorized snapshots from the simulation, with each entry of a fixed column corresponding to a QoI evaluated at a given point in space.}
      \end{figure}

We begin our background on the matrix ID by introducing conventions used throughout this work.
We assume that a data matrix $\bm{A} \in \mathbb{R}^{m \times n}$ is arranged such that each of its columns corresponds to a physical quantity of interest (QoI), such as pressure or velocity measured on a grid of size $m$. The QoI is obtained on a discrete set of locations in the physical domain, possibly without specific ordering, e.g., a flattened representation of an unstructured grid. The index of a column is assumed to correspond to a specific time-instance in the simulation from which the data is generated, as shown in Figure~\ref{fig:PDE_data}. Corresponding notation conventions used in this work are also provided in Table~\ref{tab:TableOfNotationForMyResearch}.
\alec{Importantly, for large-scale applications we never form the matrix $\bm{A}$ explicitly, as we assume that $\bm{A}$ is too large to be stored in working memory; however, we will continue to refer to $\bm{A}$ throughout the paper for clarity.}

If the spatial or temporal variation of the QoI is smooth, the matrix $\bm{A}$ may admit a low-rank structure; its columns may contain redundant information.
An assumption crucial to this work is that $\bm{A}$ is {\it numerically low rank}; i.e., its singular value decay is sharp and $\bm{A}$ may therefore be approximated with \alec{small} reconstruction error. Equivalently, the QoI may be accurately reconstructed on a low-dimensional linear subspace.

\begin{table}[htbp]
\centering 
\begin{tabular}{r c p{10cm} }
$\bm{A}$ & $\triangleq$ & data matrix\\
$\bm{B}$ & $\triangleq$ & data matrix sketch\\
$\bm{C}$ & $\triangleq$ & column ID coefficient matrix \\
$\bm{M}$ & $\triangleq$ & interpolation operator \\
$\bm{a}_i$ & $\triangleq$ & $i^{th}$ column of $\bm{A}$, i.e., $\bm{A}(:,i-1)$\\
$\mathcal{I}$ & $\triangleq$ & column index vector\\
$m$ & $\triangleq$ & size of fine grid mesh \\
$m_c$ & $\triangleq$ & size of coarse grid mesh \\
$n$ & $\triangleq$ & number of snapshots  \\
$t$ & $\triangleq$ & number of points in stencil \\
$\bm{A}(:,\mathcal{I})$ & $\triangleq$ & column skeleton of data matrix\\ 
$\bm{A}(\mathcal{J},:)$ & $\triangleq$ & row skeleton of data matrix\\
$\bm{A}(:,:k-1)$ & $\triangleq$ & first $k$ columns of $\bm{A}$ \\ 
$\bm{A}(:,k:)$ & $\triangleq$ & last $n-k$ columns of $\bm{A}$ \\ 
$+$ & $\triangleq$ & Moore-Penrose pseudo-inverse \\
$\Vert \cdot \Vert_2$ & $\triangleq$ & Matrix spectral norm\\
$\Vert \cdot \Vert_F$ & $\triangleq$ & Matrix Frobenius norm\\
\end{tabular}
\caption{Notation conventions used in this work.}
\label{tab:TableOfNotationForMyResearch}
\end{table}

We assume that there are $k$ columns of $\bm{A}$ comprising this low-dimensional linear subspace, and that $k$ is much smaller than the original column dimension $n$. This assumption may be rephrased as follows. For each column of $\bm{A}$, which we denote $\bm{a}_i = \bm{A}(:,i-1)$, $k \ll n$, with $\bm{v}_i$ such that $\Vert \bm{v}_i \Vert_2 = 1$, and $\epsilon$ such that $0 < \epsilon/\Vert \bm{a}_i \Vert_2 \ll 1$, we may write 
\begin{equation}
    \bm{a}_i = \sum_{j=0}^{k-1} c_{ij} \bm{a}_{\mathcal{I}(j)} + \epsilon \bm{v}_i ,
\end{equation}
where $\mathcal{I}\subseteq\{0,\dots,n-1\}$ and $\vert \mathcal{I} \vert = k$. In words, any column of $\bm{A}$ may be approximated by a linear combination of a fixed subset of $k$ of its columns, indexed by $\mathcal{I}$. This constitutes a $k$-rank approximation to the matrix $\bm{A}$, where $k$ is the {\it target rank} of the approximation.

The coefficients $c_{ij}$, comprising the vector $\bm{c}_i$, are obtained via the least squares problem, 
\begin{equation}
    \bm{c}_i = \argmin_{\hat{\bm{c}}} \Vert \bm{a}_i - \sum_{j=0}^{k-1} \hat{\bm{c}}_{j} \bm{a}_{\mathcal{I}(j)} \Vert_2 .
\end{equation}

The approximations of each column $\bm{a}_i$ may be concatenated to form an approximation to the original data matrix $\bm{A}$ via the matrix-matrix product
\begin{equation}
    \label{eq:columnID}
    \bm{A} \approx \bm{A}(:,\mathcal{I})\bm{C} , 
\end{equation}
where $\bm{C} \in \mathbb{R}^{k \times n}$ is the {\it coefficient matrix} whose entries are comprised of the least-squares coefficients $c_{ij}$ corresponding to  each column $\bm{a}_i$:
\alec{\begin{equation}
    \bm{C} = \begin{bmatrix}
   \bm{c}_0 & \bm{c}_1 & \cdots & \bm{c}_{k-2} & \bm{c}_{k-1} \\
   \end{bmatrix} .
\end{equation}}

The matrix $\bm{A}(:,\mathcal{I}) \in \mathbb{R}^{m \times k}$, which is a subset of the columns of $\bm{A}$ indexed by $\mathcal{I}$, is called the {\it column skeleton} of $\bm{A}$. The decomposition (\ref{eq:columnID}) is the so-called column interpolative decomposition (column ID) of a matrix $\bm{A}$~\cite{cheng2005compression}. For the interested reader, properties of the column ID are enumerated in Lemma~\ref{lem:IDproplemma} in the Appendix.

The column ID possesses favorable properties relative to orthogonal decompositions such as the SVD, among them~\cite{martinsson2019randomized}:
\begin{itemize}
\item It yields factor matrices which are sparsity and \alec{entry-wise} non-negativity preserving; if $\bm{A}$ is sparse or \alec{entry-wise} non-negative, so is $\bm{A}(:,\mathcal{I})$.
\item The column ID is self-expressive; the matrix is decomposed in terms of one of its own \alec{columns}. For domain scientists, this may make the factor matrices in the ID more \alec{{\it interpretable}} than, e.g, the singular value decomposition (SVD).
\item \alec{The column skeleton extracted by column ID preserves physical constraints from the original data matrix.}
    \end{itemize}

There are numerous schemes to produce a column ID; in this work, we deploy the column pivoted QR based on a modified Gram-Schmidt scheme~\cite{golub2012matrix}. Approaches such as the rank-revealing QR~\cite{gu1996efficient} provide stronger theoretical guarantees than a basic pivoted QR, though these guarantees come at the cost of increased computational expense. Other methods available for computing decompositions comprised of a column subset and corresponding coefficient matrix may be found in, e.g.,  \cite{mahoney2009cur,elhamifar2009sparse,dyer2015self,perry2019allocation}. 

In order to form the ID, we first form a column pivoted QR of $\bm{A}$:

\begin{equation}
\bm{A}\bm{Z} = \bm{Q}\bm{R} ,
\end{equation}
where $\bm{Z}$ is a permutation encoding the pivots, $\bm{Q} \in \mathbb{R}^{m \times n}$ is a unitary matrix, and $\bm{R} \in \mathbb{R}^{n \times n}$ is an upper triangular matrix. We then partition $\bm{Q}$ and $\bm{R}$ into blocks such that
\begin{equation}
\bm{Q} =  \begin{bmatrix} \bm{Q}_{11} & \bm{Q}_{12} \end{bmatrix}  , 
\end{equation}
where $\bm{Q}_{11} \in \mathbb{R}^{m \times k}$ and $\bm{Q}_{12} \in \mathbb{R}^{m \times (n-k)}$, and
\begin{equation}
\bm{R} = \begin{bmatrix}
            \bm{R}_{11} & \bm{R}_{12} \\
        \bm{0}  & \bm{R}_{22}
     \end{bmatrix} , 
\end{equation}
where $\bm{R}_{11} \in \mathbb{R}^{k \times k}$, $\bm{R}_{12} \in \mathbb{R}^{k \times (n-k)}$, and $\bm{R}_{22} \in \mathbb{R}^{(n-k) \times (n-k)}$. We then form a low-rank approximation of $\bm{A}$:
\begin{align}
\bm{A}\bm{Z} &\approx \bm{Q}_{11} \left[ \bm{R}_{11} \hspace{6pt} \bm{R}_{12} \right] ,
\\
&= \bm{Q}_{11}\bm{R}_{11} \left[ \bm{I} \hspace{6pt} \bm{R}_{11}^{+} \bm{R}_{12} \right] .
\end{align} 
As $\bm{Q}_{11}\bm{R}_{11} = \bm{A}\bm{Z}(:,\mathcal{I})$, it follows that 
\begin{equation}
\bm{A} \approx \bm{A}(:,\mathcal{I}) \left[ \bm{I} \hspace{6pt} \bm{R}_{11}^{+} \bm{R}_{12} \right] \bm{Z}^T .
\end{equation} 
Setting 
\begin{equation}
\bm{C} = \left[ \bm{I} \hspace{6pt} \bm{R}_{11}^{+} \bm{R}_{12} \right] \bm{Z}^T ,
\end{equation}
we obtain a column ID of $\bm{A}$:
\begin{equation}
    \bm{A} \approx \bm{A}(:,\mathcal{I})\bm{C} \label{eqn:fullID}.
\end{equation}

We present the Column ID as Algorithm~\ref{alg:id}. Our approach takes as input an approximation rank in Step 2, then computes a rank-$k$ pivoted QR decomposition of the input matrix in Step 3 \alec{based on a modified Gram-Schmidt procedure  -- denoted {\it mgsqr} within Algorithm~\ref{alg:id} -- see~\cite{golub2012matrix} for more details}. Steps 4 through 6 demonstrate the formation of the coefficient matrix $\bm{C}$. This, combined with the index vector obtained in Step 3, enables construction of the final approximation of our matrix.

\begin{algorithm}[h]
\caption{Column ID $\bm{A} \approx \bm{A}(:,\mathcal{I})\bm{C}$~\citep{cheng2005compression}}	\label{alg:id}
\begin{algorithmic}[1]
\Procedure{ID}{$\bm{A}$ $\in \mathbb{R}^{m \times n}$}
\State $k \gets$ approximation rank 
\State $\bm{Q}$, $\bm{R}$, $\mathcal{I} \gets mgsqr(\bm{A},k)$
\alec{\State $\bm{R}_{11} \gets \bm{R}(:k-1,:k-1)$
\State $\bm{R}_{12} \gets \bm{R}(:k-1,k:)$ 
\State $\bm{Z} \gets \bm{I}_n(:,[\mathcal{I}\ ,\ \mathcal{I}^c])$ \hspace{4.4cm} 
\State $\bm{C} \gets \left[\bm{I}_k \hspace{2pt}\vert\hspace{2pt}\bm{R}_{11}^+\bm{R}_{12} \right]\bm{Z}^{T}$}
\State $\bm{return}$ $\bm{A}(:,\mathcal{I})$, $\bm{C}$  \hspace{4.5cm}
\EndProcedure
\end{algorithmic}
\end{algorithm}

\alec{Computing the ID using a pivoted QR-based approach costs at most $\mathcal{O}(mn\min(m,n))$ FLOPs; the cost is typically closer to $\mathcal{O}(mnk)$.}
Several strategies are available to reduce the computational cost of computing the column ID of $\bm{A}$. Some schemes are based on the observation that the column indices $\mathcal{I}$ and coefficient matrix $\bm{C}$ may be computed on a lower dimensional matrix {\it sketch} of $\bm{A}$, herein denoted $\bm{B} \in \mathbb{R}^{\ell \times n}$ with $k < \ell \ll m$. Using the sketch to form the ID works if $\bm{B}$ is constructed from $\bm{A}$ such that it encodes a sufficient amount of geometric information \alec{from the range of the original matrix $\bm{A}$. More precisely, let $\bm{B}^+\bm{B}$ be the orthogonal projection onto the range of $\bm{B}$. Then, we may seek a sketch $\bm{B}$ which satisfies}
\begin{equation}
    \Vert \bm{A} - \bm{A}(\bm{B}^+\bm{B}) \Vert_{\xi} \leq \epsilon \Vert \bm{A} \Vert_{\xi}; \qquad \xi = 2,F.
\end{equation}

In words, if $\bm{A}$ projected onto the range of $\bm{B}$ is close in proximity to $\bm{A}$ with respect to, e.g., the spectral or Frobenius norm, then $\bm{B}$ will perform well as a matrix sketch in a column ID. 
To form a column ID using a sketch $\bm{B}$, we first compute its column ID,
\begin{equation}
\bm{B} \approx \bm{B}(:,\mathcal{I}_B)\bm{C}_B.
\end{equation}
$\mathcal{I}_B$ and $\bm{C}_B$ may then be used to construct a column ID for $\bm{A}$,
\begin{equation}
\bm{A} \approx \bm{A}(:,\mathcal{I}_B)\bm{C}_B .
\end{equation}
The cost of computing the index vector $\mathcal{I}_B$ and coefficient matrix $\bm{C}_B$ is reduced significantly relative to computing $\mathcal{I}$ and $\bm{C}$ from the full matrix as in (\ref{eqn:fullID}). Using $\bm{B}$ in place of $\bm{A}$ also increases parallel efficiency in ID algorithms, as we demonstrate in Section~\ref{sec:numerical_experiments}.

Forming the matrix $\bm{B}$ has been an active area of research for the past few decades. Many approaches use random linear embeddings to map data into lower dimensions, referred to as {\it random projections}. In one example of random projection, the matrix $\bm{A}$ is left-multiplied by a matrix $\bm{\Omega} \in \mathbb{R}^{\ell \times m}$ whose entries are independent and identically distributed Gaussian random variables. This generates the sketch matrix $\bm{B} = \bm{\Omega}\bm{A}$~\cite{liberty2007randomized}. For a thorough review on the subject, we refer the interested reader to~\cite{woodruff2014sketching}.

Following~\cite{dunton2020pass}, this work takes a  different approach, focusing on {\it deterministically} projecting $\bm{A}$ to a lower dimension by constructing a {\it coarse grid representation} which exploits the smoothness of the QoI in the spatial domain, as opposed to the data-agnostic nature of randomized sketches. One deterministic strategy entails subsampling the \alec{rows} of $\bm{A}$, and therefore the domain of the corresponding QoI, to obtain the sketch matrix \alec{$\bm{B}$}.
Computing the column ID,
\begin{equation}
\label{eq:coarseA_ID}
  \bm{B} \approx \bm{B}(:,\mathcal{I}_B)\bm{C}_B = \bm{A}(\mathcal{J},\mathcal{I}_B)\bm{C}_B  ,
\end{equation} 
yields the components required to generate a viable column ID of $\bm{A}$,
\begin{equation}
\label{eq:coarseregA}
\bm{A} \approx \bm{A}(:,\mathcal{I}_B)\bm{C}_B .
\end{equation}

Forming a column ID of $\bm{A}$ from $\bm{B} = \bm{A}(\mathcal{J},:)$ in this way constitutes the subsampled ID (SubID), herein presented as Algorithm~\ref{alg:subsampid}. The procedure is nearly identical that of the standard Column ID, except we subsample the input matrix $\bm{A}$ in Step 3 prior to the pivoted QR computation. The index vector and coefficient matrix computed from the subsampled matrix are used to form the final approximation of our matrix. Accuracy guarantees for SubID are provided in the appendix; the interested reader is referred to~\cite{dunton2020pass} for more details on this method.

\begin{algorithm}[h]
\caption{SubID $\bm{A} \approx  \bm{A}(:,\mathcal{I}_B) \bm{C}_B$~\cite{dunton2020pass}}	\label{alg:subsampid}
\begin{algorithmic}[1]
\Procedure{SubID}{$\bm{A}$ $\in \mathbb{R}^{m \times n}$}
\State $k \gets$ approximation rank 
\State $\bm{B} \gets subsample(\bm{A})$ \hspace{4.1cm}
\State $\bm{Q}_B$, $\bm{R}_B$, $\mathcal{I}_B \gets mgsqr(\bm{B},k)$
\alec{\State $\bm{R}_{11} \gets \bm{R}_B(:k-1,:k-1)$
\State $\bm{R}_{12} \gets \bm{R}_B(:k-1,k:)$ 
\State $\bm{Z}_B \gets \bm{I}_n(:,[\mathcal{I}_B\ ,\ \mathcal{I}_B^c])$ \hspace{4.4cm} 
\State $\bm{C}_B \gets \left[\bm{I}_k \hspace{2pt}\vert\hspace{2pt}\bm{R}_{11}^+\bm{R}_{12} \right]\bm{Z}_B^{T}$}
\State $\bm{return}$ $\bm{A}(:,\mathcal{I}_B)$, $\bm{C}_B$  \hspace{4.5cm}
\EndProcedure
\end{algorithmic}
\end{algorithm}

\alec{\begin{table*}[thb]
\centering
\tabcolsep7pt\begin{tabular}{lccc}
\hline
Method & Computational complexity & Disk storage & RAM usage\\
\hline
ID (Algorithm~\ref{alg:id}) & $\mathcal{O}(mnk)$  & $k(m+n)$ & $k(m+n) + mn$ \\
SubID (Algorithm~\ref{alg:subsampid}) & $\mathcal{O}(m_cnk)$ & $k(m+n)$ & $k(m_c + n) + m_cn$\\
SPID (Algorithm~\ref{alg:spid}) & $\mathcal{O}(m_cnk)$ & $k(m_c + n) + tm$ & $k(m_c+n) + m_cn + tm$\\
\hline
\end{tabular}
\caption{Computational complexity of ID, subsampled ID and single-pass ID with input matrix of size $m \times n$ and target rank $k$. Variable $m_c \ll m$ represents the row dimension of the input matrix after subsampling. $t$ is the size of the stencil for the interpolation scheme. Disk storage is given as the total number of matrix entries stored following execution of the algorithm.}	\label{tab:methods}
\end{table*}}
Computing a column ID using SubID requires two passes over the data matrix: one pass to compute the indices $\mathcal{I}$ and coefficient matrix $\bm{C}$, then a second pass to obtain the column skeleton $\bm{A}(:,\mathcal{I})$. When the amount of data generated in the solver step (which generates the data matrix) exceeds the storage available in RAM or disk, the user may be required to run a second simulation in order to construct $\bm{A}(:,\mathcal{I})$. In order to avoid this second pass over $\bm{A}$ -- and potentially the need to run a second simulation -- we also present a single-pass variation of SubID: the so-called single-pass ID (SPID)~\cite{dunton2020pass}. In this variation of column ID, instead of generating the approximation $\bm{A} \approx \bm{A}(:,\mathcal{I}_B)\bm{C}_B$ as in SubID, we approximate $\bm{A}$ as
\begin{equation}
    \bm{A} \approx \bm{M}\bm{B}(:,\mathcal{I}_B)\bm{C}_B ,
    \label{eq:single_ID}
\end{equation}
where $\bm{M}$ is an interpolation operator which interpolates the coarse grid QoI matrix $\bm{B}$ onto the fine grid. Accuracy guarantees for SPID are provided in the appendix; the interested reader is also referred to~\cite{dunton2020pass} for more details.

We present SPID as Algorithm~\ref{alg:spid}. This algorithm only differs from SubID in that in the final step, instead of returning $\bm{A}(:,\mathcal{I}_B)$, we return the column skeleton of the coarsened data matrix $\bm{B}(:,\mathcal{I}_B)$ and the interpolation operator $\bm{M}$. Both the column indices $\mathcal{I}_B$ and coefficient matrix $\bm{C}_B$ are computed from the coarse grid data matrix $\bm{B}$.

\begin{algorithm}[h]
\caption{SPID $\bm{A} \approx \bm{M}\bm{B}(:,\mathcal{I}_B) \bm{C}_B$~\cite{dunton2020pass}}	\label{alg:spid}
\begin{algorithmic}[1]
\Procedure{SPID}{$\bm{A}$ $\in \mathbb{R}^{m \times n}$}
\State $k \gets$ approximation rank 
\State $\bm{B} \gets subsample(\bm{A})$ \hspace{4.1cm}
\State $\bm{Q}_B$, $\bm{R}_B$, $\mathcal{I}_B \gets mgsqr(\bm{B},k)$
\alec{\State $\bm{R}_{11} \gets \bm{R}_B(:k-1,:k-1)$
\State $\bm{R}_{12} \gets \bm{R}_B(:k-1,k:)$ 
\State $\bm{Z}_B \gets \bm{I}_n(:,[\mathcal{I}_B\ ,\ \mathcal{I}_B^c])$ \hspace{4.4cm} 
\State $\bm{C}_B \gets \left[\bm{I}_k \hspace{2pt}\vert\hspace{2pt}\bm{R}_{11}^+\bm{R}_{12} \right]\bm{Z}_B^{T}$}
\State $\bm{return}$ $\bm{B}(:,\mathcal{I}_B)$, $\bm{C}_B$, $\bm{M}$ \hspace{4.5cm}
\EndProcedure
\end{algorithmic}
\end{algorithm}

The column ID enables data compression due to the fact that the factor matrices are comprised of a total of $k(m +n)$ entries -- in the case of SPID this is even smaller -- while the original matrix $\bm{A}$ contains $mn$ elements. Under the assumption that $\bm{A}$ is numerically low rank, we may achieve an accurate column ID low-rank approximation with $k \ll m$ and $k \ll n$, and thus $k(m+n) \ll mn$. Therefore, if $\bm{A}$ is numerically low-rank, storing $\bm{A}(:,\mathcal{I})$ and $\bm{C}$ in lieu of $\bm{A}$ will yield significant memory savings. We measure these savings using a metric called the {\it compression factor} (CF), which we define
\begin{equation} \label{cf}
    \text{CF} \vcentcolon= \frac{mn}{k(m+n)}.
\end{equation}

\alec{We provide results for computational complexity, disk storage, and RAM usage for the three algorithms in Table~\ref{tab:methods}. We observe that SubID and SPID both offer improvements over ID in terms of complexity, reducing the complexity from $\mathcal{O}(mnk)$ to $\mathcal{O}(m_c nk)$ with $m_c \ll m$. SPID also decreases the amount of memory used to store the compressed data to disk. In terms of RAM usage, SubID is slightly more efficient than SPID due to the interpolation step required in SPID. }

\begin{remark}
Although we have used the convention that the rank of the ID approximation is fixed to be $k$ for simplicity, the {\it mgsqr} subroutine may be augmented to adaptively determine the rank of the approximation. Once the approximation reaches a given tolerance, the method terminates and returns a target rank as output, along with a corresponding column skeleton and coefficient matrix.
\end{remark}

\section{ID methods and task-based parallelism}
\label{sec:ID_task_based}
 In this section, we outline the implementation of SubID and SPID as compressors of simulation data in a task-based parallel environment. We begin with an overview of parallelism in CFD codes, followed by parallelism in ID codes, then provide a description of how an ID routine may be embedded in a CFD code, and finally a detailed description of the implementation.
\subsection{Parallelism of CFD codes}

\begin{figure}[!h]
       \center{\includegraphics[width=0.49\textwidth]
       {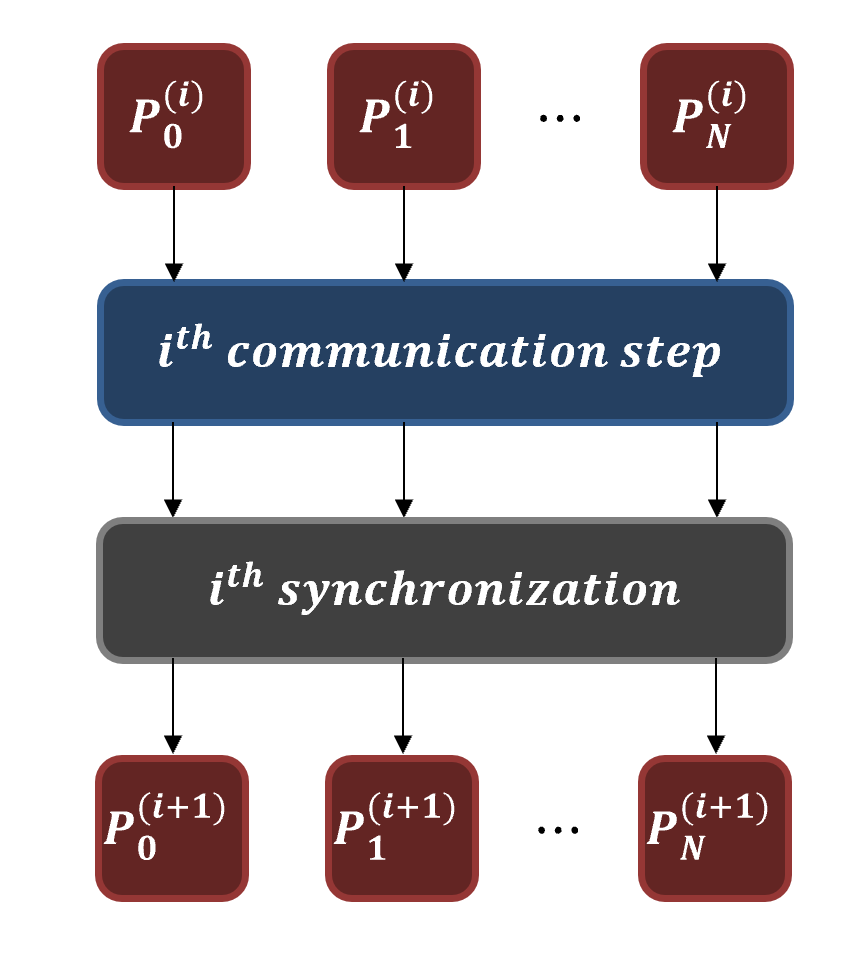}}
       \caption{\label{fig:CFD_flow} Flowchart for a general CFD process; processors $P_0$, $P_1$, ..., $P_N$ at the $i^{th}$ time step perform required calculations, exchange data, then synchronize before continuing on to the next time step.}
      \end{figure}

Computational fluid dynamics (CFD) simulations are concerned with the numerical solution of a set of partial differential equations (PDEs) on a discretized grid representation of the physical domain of interest. For large simulations, this discretized domain is partitioned into sub-domains (blocks) of data, which are then distributed across the available computing resources in a process called domain decomposition. This strategy of extracting concurrency from a given computational approach is referred to as data-parallelism. Each ``step" of the simulation consists of two coupled processes. First, identical sets of instructions are executed on each block (independent of the other blocks). Second, a communication stage, where relevant data is exchanged between the blocks of the domain (e.g., information from stencil points for a finite difference method), is performed. This ``back-and-forth" repeats until the simulation is completed. Often, these simulations are implemented using a \textit{synchronous} or \textit{bulk-synchronous} programming model; that is, a barrier is inserted after each step of the simulation to prevent the simulation from proceeding until all current processes are complete \alec{and have exchanged the information required to neighboring domains to continue the computations} (see Figure~\ref{fig:CFD_flow}).

\subsection{Parallelism of ID methods}
      
From a computational perspective, ID differs greatly from CFD simulations. ID compression algorithms can be applied independently to blocks of arbitrary sizes; no communication is required between these blocks to perform the data compression. This means that the user does not need to devise a separate domain decomposition strategy for the ID; the existing flow solver decomposition is suitable. Additionally, because there is no communication involved between the sub-domains, the ID compression can begin as soon as all required data has been computed, without the need for a synchronization stage. So, if the ID compression is implemented in such a way that it begins while the flow solver is still running, the computations required to perform the ID analysis on the most recent solution time step can be performed while waiting for the data synchronization required to compute the flow solver update for the next simulation step. Thus, the overhead cost of performing the ID approximation can be reduced or hidden if the computing system has a large amount of communication latency (as is often the case for extremely large simulations).

\begin{figure}[!h]
       \center{\includegraphics[width=0.49\textwidth]
       {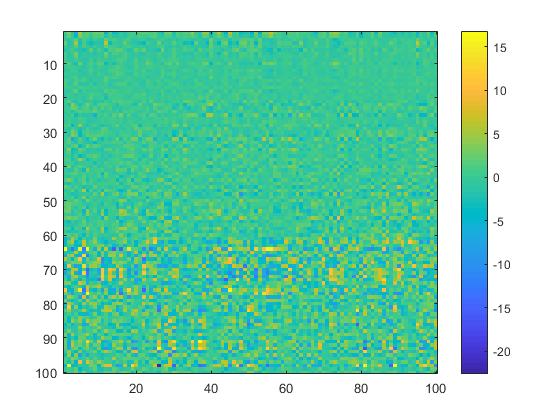}}
       \caption{\label{fig:full_matrix} Low-rank Test Matrix}
       \end{figure}

 In addition to increasing computational efficiency, applying ID individually to sub-domains of the data can help to increase the overall compression. This is due to the fact that \alec{the data corresponding to some of the subdomains} may be locally low-rank, i.e., on individual sub-blocks, but globally high-rank.
 As an example, consider Figure~\ref{fig:full_matrix}, whose color map displays the values of the entries of a simple, low-rank test matrix.

\begin{figure}[!h]
       \center{\includegraphics[width=0.49\textwidth]
       {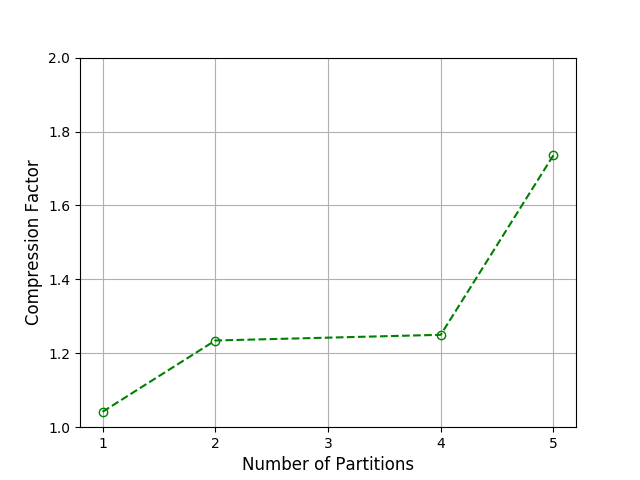}}
       \caption{\label{fig:cf_test_matrix} Matrix compression factor \heather{(Eqn.~\ref{cf})} plotted against the number of partitions of the $100 \times 100$ test matrix.}
       \end{figure}

   This test matrix, which has a rank of 48, was partitioned into various submatrices along the rows (representing \alec{typical} partitions in the spatial domain), and the ID algorithm was applied to each individual partition with a target rank of 50. The overall compression factor was then computed. 
   \heather{Table~\ref{matrix_ranks}} shows the rank of the resulting ID approximations for each partition, and 
   Figure~\ref{fig:cf_test_matrix} shows their corresponding compression factors. As the figures demonstrate, partitioning this matrix and independently applying ID to these submatrices increased the overall compression factor; when the matrix is partitioned into 5 submatrices, the overall ID compression factor is almost 70\% larger than that of the single partition. Though these results are for a specific test matrix, this behavior can be generally observed in CFD applications due to the structure of the relevant data matrices. In these matrices, adjacent row entries correspond to adjacent points in the spatial domain, and adjacent column entries correspond to adjacent points in the temporal domain. Because of this, variation of physical QoIs (whose values make up the entries of the matrix) is less drastic over local regions of the matrix. Therefore, partitioning the data matrix is a favorable approach because it allows the column ID to extract low rank structure more efficiently.
   
\begin{table}[!htb]
\begin{center}
\begin{tabular}{ |c|c| } 
\hline
No. of Partitions & Ranks of Submatrices\\
 \hline
 $1$ & $\left[48\right]$ \\ 
 \hline
 $2$ & $\left[ 24, 30 \right]$ \\
 \hline
 $4$ & $\left[ 11, 18, 18, 17 \right]$ \\
 \hline
 $5$ & $\left[ 6, 12, 6, 12, 12 \right]$ \\
 \hline
\end{tabular}
\end{center}
\caption{\heather{Rank of submatrices as a function of number of partitions of the original test matrix (Figure~\ref{fig:full_matrix}). The rank of the submatrices decreases as the number of partitions is increased.}}
\label{matrix_ranks} 
\end{table}
      
Additionally, if the QoI across the domain is highly variable, the number of calculations required to find the ID low-rank approximation for each block can vary by a large amount when using a fixed error tolerance setting instead of a fixed rank. For both of these reasons, implementing ID within a synchronous or barrier-synchronous parallelization model (such as  OpenMP, MPI, and CUDA) will lead to large reductions in performance as the simulation becomes increasingly complex and/or large. Instead, we use an \textit{asynchronous} parallelization model that removes these barriers, and select a programming model that is well-suited to simulations that are run on modern, heterogeneous supercomputers.

\subsection{Task-based Parallelization of ID methods in CFD codes} 

\begin{figure}[!h]
       \center{\includegraphics[width=0.49\textwidth]
       {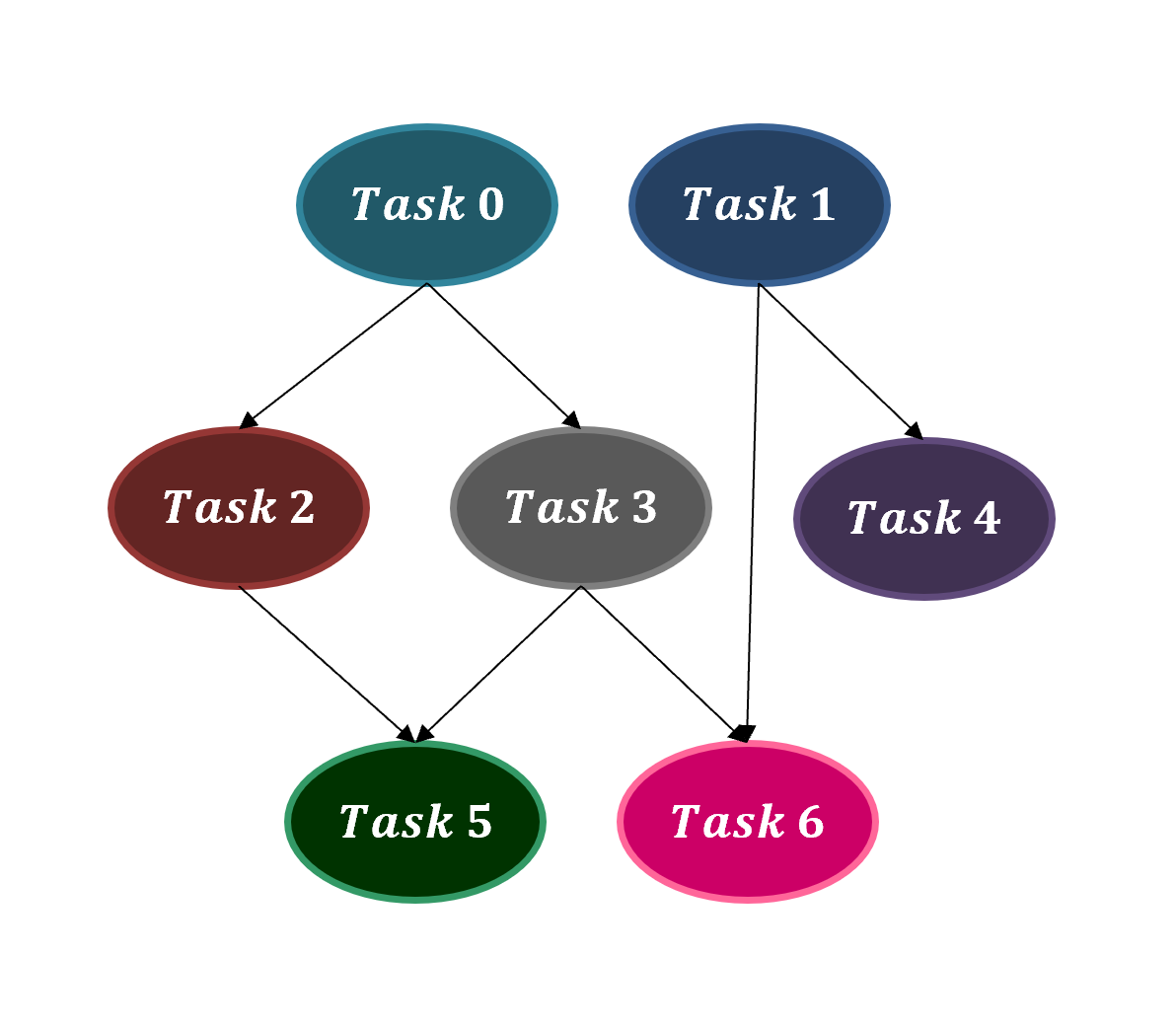}}
       \caption{\label{fig:task_graph} Task graph for task-based parallelism.}
      \end{figure}

Task-based (or functional) parallelism is one such model that meets these requirements. As the name implies, the parallelism is based on tasks, which are units of work made up of statements, blocks, or loops within the program. These tasks can operate on independent or dependent subsets of data; the data dependencies between the tasks of a program can be represented as a task graph, as shown in Figure~\ref{fig:task_graph}. Each node of the graph represents an individual task and each edge represents a dependency. Parallel execution of the program is opportunistic; it is only dependent upon information contained in the task graph, with the launch-time of a task determined by any dependencies on previously launched tasks. There are no barriers in the program aside from these task dependencies, making program execution asynchronous \cite{parallel_text}.

There are \alec{several} advantages to using task-based parallelism, among other parallel programming models, for extremely large and complex simulations. Scientific computing is increasingly trending toward multi-physics, multi-scale applications. A task-based approach is well-suited for complex codes, where sub-domains of the data might not necessarily require the same operations. Additionally, the current trend in computer architectures is the adoption of accelerator and, in general, heterogeneous computing frameworks with complex memory hierarchies.   Task-based parallelism allows the user to take advantage of hybrid computer architectures because it opportunistically assigns tasks to the available ``worker pool" of compute resources, removing the need for user-specified load balancing for each type of processor \heather{\cite{parallel_text}}. 

Task-based parallelism also reduces synchronization bottlenecks experienced by data-based parallelism on large machines, which is crucial because the costs of data movement within these architectures dominates overall compute time as the cost of floating point operations exponentially decreases.  Thus, task-based parallelism provides an ideal environment for the implementation of ID algorithms within CFD codes. One example of a programming system that uses this model is Legion~\cite{Legion_paper}, which will be discussed in more detail in Section~\ref{sec:legion}.

\subsection{Implementation} \label{id_imp}

\begin{figure}[!h]
   \center{\includegraphics[width=0.49\textwidth]
   {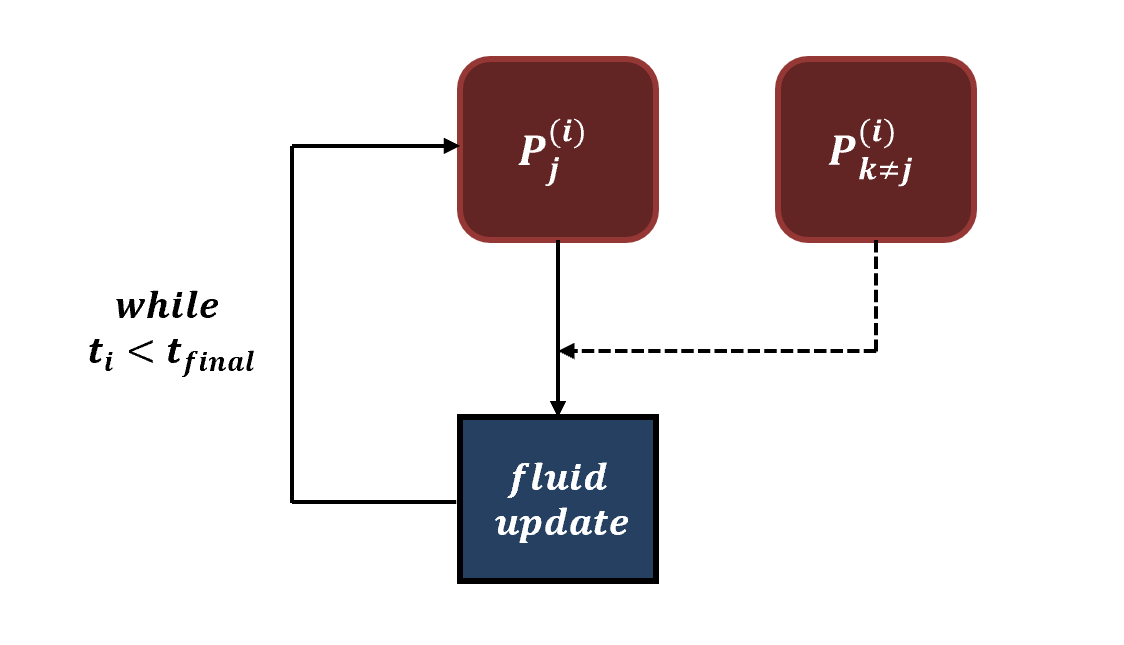}}
   \caption{\label{fig:fluid_flowchart} Schematic of a task-parallel CFD code.}
  \end{figure}

A task-parallel CFD code will have a layout like that shown in Figure~\ref{fig:fluid_flowchart}. Each processor contains local data about a subsection of the physical domain. At the $i^{th}$ time step of the simulation, the $j^{th}$ processor $P_j$ receives data from the other compute processors, $P_{k \neq j}$, which is used in addition to its local data to perform computations and advance the fluid solver in time. \alec{$P_j$ sends information to all of the other processors as well. However, this send message is non-blocking for $P_j$.} This process repeats until the final simulation time is reached.

  As shown in Algorithm~\ref{alg:cfd}, this can be represented as a for-loop nested within a while-loop. The task \textit{update\_fluid\_step} is launched on all $N+1$ processors at each time step of the simulation. This task requires both the local processor's block of fluid domain data, as well as the data from the stencil regions of the block. As expected, there are no synchronization barriers in this implementation; dependencies in execution time depend only upon the communication of the stencil data between processors. \alec{This data exchange is to set the boundary conditions in each subdomain; the stencil determines how these subdomains are connected.}

\begin{algorithm}[h]
\caption{Task-Parallel CFD Solver}	
\label{alg:cfd}
\begin{algorithmic}[1]
\While{$t < t_{final}$}
\For{$\text{id} = 0, N$}  
  \State $\text{update\_fluid\_step(block[id], stencil[id]})$
\EndFor
\EndWhile
\end{algorithmic}
\end{algorithm}

\begin{figure}[!h]
   \center{\includegraphics[width=0.49\textwidth]
   {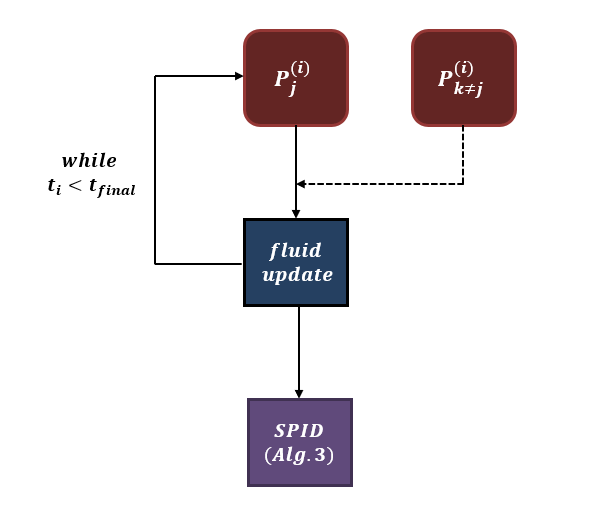}}
   \caption{\label{fig:fluid_naive_flowchart} \heather{Application of naive SPID (Algorithm~\ref{alg:spid}) to a CFD simulation.}}
  \end{figure}

\heather{\subsubsection{Naive SPID}}
The SPID algorithm detailed in Section 2 is performed after the completion of the flow solver simulation. Figure~\ref{fig:fluid_naive_flowchart} shows this application execution. In this implementation, the entirety of the flow simulation data (that is, the values of the QoI in the fluid domain at each time step) must be stored. These are then used for SPID.

\begin{algorithm}[h]
\caption{Task-Parallel CFD Solver with Naive SPID}	
\label{alg:cfd_w_SPID}
\begin{algorithmic}[1]
\While{$t < t_{final}$}
\For{$\text{id} = 0, N$}  
  \State $\text{update\_fluid\_step(block[id], stencil[id]})$
  \State $\text{store\_subsampled\_data(block[id])}$
\EndFor
\EndWhile
\For{$\text{id} = 0, N$}  
  \State $\text{SPID(subsampled\_block[id])}$
\EndFor
\end{algorithmic}
\end{algorithm}

\begin{figure}[!h]
       \center{\includegraphics[width=0.49\textwidth]
       {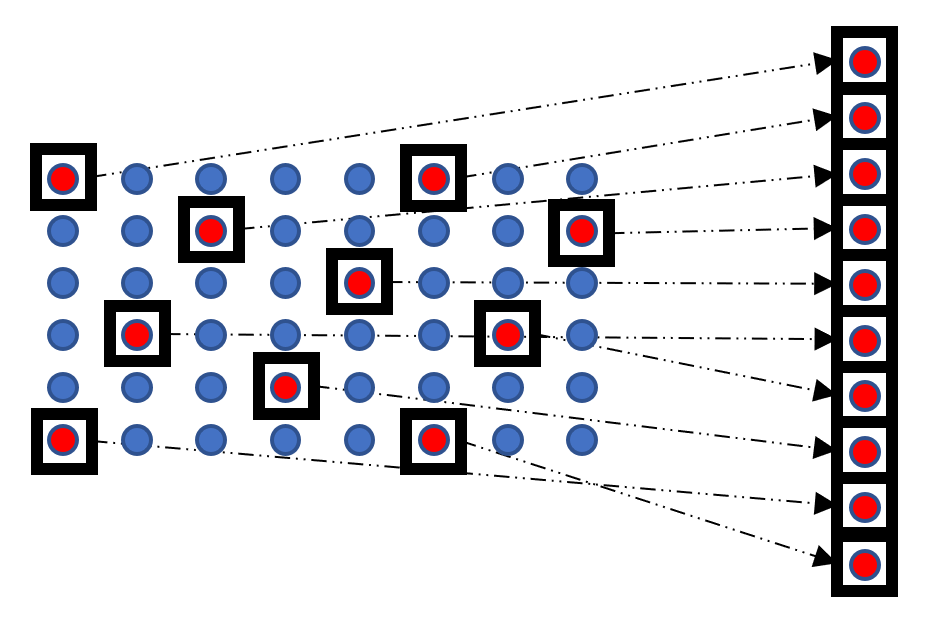}}
       \caption{\label{fig:sketch_diagram} Formation of column vector of sketch matrix with every $5^{th}$ entry subsampled on a 2D domain.}
      \end{figure}

Algorithm~\ref{alg:cfd_w_SPID} shows the additional steps required to perform the SPID. Each processor must again perform the fluid update at each time step, but now the data must also be stored, which requires the addition of the \textit{store\_subsampled\_data} task. This data, an instantaneous temporal snapshot of the domain, is flattened and stored in a column of the data matrix $\bm{A}$. This flattening process can be seen in Figure~\ref{fig:sketch_diagram}, where we display the formation of a column of a subsampled sketch of the matrix $\bm{A}$. Selecting every $5^{th}$ grid point on a 2D spatial domain, we are able to embed data collected from a mesh with $48$ degrees of freedom into a $10$-dimensional space. This type of embedding not only reduces the dimension of our data (and consequently its memory footprint), but exploits smoothness in the spatial domain\heather{, when it exists}.

After the fluid solve has completed, the \textit{SPID} task is launched (in parallel) across all of the processors. This task requires performing a pivoted QR on the entire data matrix $\bm{A}$. If the number of time steps, $n$, is large, e.g, on the order of the number required for an explicit time-advancement scheme, then the amount of memory required to store and perform ID on the fluid domain greatly exceeds what is available, even on large distributed-memory systems. To circumvent this problem, we apply the ID algorithm in a \alec{two-stage manner}; \heather{this implementation is detailed in the following section}. 

\heather{\subsubsection{Modified SPID} \label{modified_spid}} 
\heather{This two-stage algorithm, referred to as Modified SPID, reduces the size of the data matrix being operated on and allows the data compression algorithm to begin while the flow solver is still running.} \\

\heather{\textbf{Stage 1:}} The first stage applies the ID algorithm \heather{with a target rank of $k$} to $N$ individual blocks of time steps of the domain, so that the following matrix, which we denote $\bm{A^0}$, is formed: 

\begin{align}
    \bm{A^0} = [ \bm{A_0}(:,\mathcal{I_\text{0}})\bm{C_0} \cdots  \bm{A_{N-1}}(:,\mathcal{I_{N-\text{1}}})\bm{C_{N-1}} ] ,
\end{align}

\noindent where \alec{the subscript} $0$ corresponds to the first $\frac{n}{N}$ time steps, index $1$ corresponds to the next $\frac{n}{N}$ time steps, and so on until index $N-1$. 

We may then decompose $\bm{A^0}$ to be the product
\begin{align}
    \bm{A^0} & = \bm{A^{0'}} \bm{C^{0'}} \\ &= [ \bm{A_0}(:,\mathcal{I_\text{0}})  
 \cdots \bm{A_{N-1}}(:,\mathcal{I_{N-\text{1}}}) ] \bm{C^{0'}}  , 
\end{align}
where $\bm{C^{0'}} \in \mathbb{R}^{k(N-1) \times N}$ is a sparse matrix comprised of rectangular blocks given coefficient matrices $\bm{C}_i$, constructed as follows:

\begin{gather}
 \bm{C^{0'}}
 =
  \begin{bmatrix}
   \bm{C}_0 & \bm{0} & \bm{0} & \cdots & \bm{0} \\
   \bm{0} & \bm{C}_1 & \bm{0} & \cdots & \bm{0} \\
   \bm{0} & \bm{0} & \bm{C}_2 &  \cdots & \bm{0} \\
   \vdots & \vdots  & \vdots &  \ddots & \vdots \\
   \bm{0} & \bm{0} & \bm{0} &  \cdots & \bm{C}_{N-1} \\
   \end{bmatrix} .
\end{gather}

\heather{\textbf{Stage 2:}} To eliminate any redundant columns in the horizontally concatenated column skeleton matrix $\bm{A^{0'}}$ and achieve further compression, we apply the ID algorithm, \heather{this time with a tolerance instead of a target rank,} to $\bm{A^{0'}}$ to obtain the following approximation:
\begin{equation}
    \bm{A^{0'}} \approx \bm{A^{1'}} \bm{C^{1}} ,
\end{equation}
where  
\begin{align}
  \bm{A^{0'}} &= \bm{A^0}(:,\mathcal{I}_U) , \\
\alec{  \mathcal{I}_U} &= \bigcup_{j=0}^{n-1} \left(\mathcal{I}_j + \frac{jn}{N} \right), \\
\bm{A^{0'}} &\approx \bm{A^{1'}}\bm{C^{1}} .
\end{align}

$\mathcal{I}_U$ indexes the columns of the original matrix $\bm{A^0}$ extracted prior to second application of column ID, which yields $\bm{A^{1'}}$ and $\bm{C^1}$. The subscript $U$ suggests that these indices come from the union of the index vectors $\mathcal{I}_j$ in $\bm{A^{0'}}$ (with an added term to adjust for the indexing in the original data stream). A similar approach to this is employed in~\cite{pi2013scalable}, though their approach involves a truncated SVD as a pre-processing step, \heather{and does not} emphasize pass-efficiency.

\heather{The} final ID approximation of the entire data is
\begin{equation}
    \bm{A} = \bm{A^{0'}} \bm{C^{0'}} \approx \bm{A^{1'}} \bm{C^{1}} \bm{C^{0'}} = \bm{A^{1'}} \bm{C^{1'}} \heather{,}
\end{equation}
\heather{where} $\bm{A^{1'}} \in \mathbb{R}^{m \times k}$ and $\bm{C^{1'}} \in \mathbb{R}^{k \times n}$ .

\begin{remark}
\alec{The same procedure may be applied} to \heather{the} case in which we block along the spatial domain; we would simply just replace $\bm{A}$ with $\bm{A}^T$ in the above analysis.
\end{remark}

\begin{figure}[!h]
   \center{\includegraphics[width=0.49\textwidth]
   {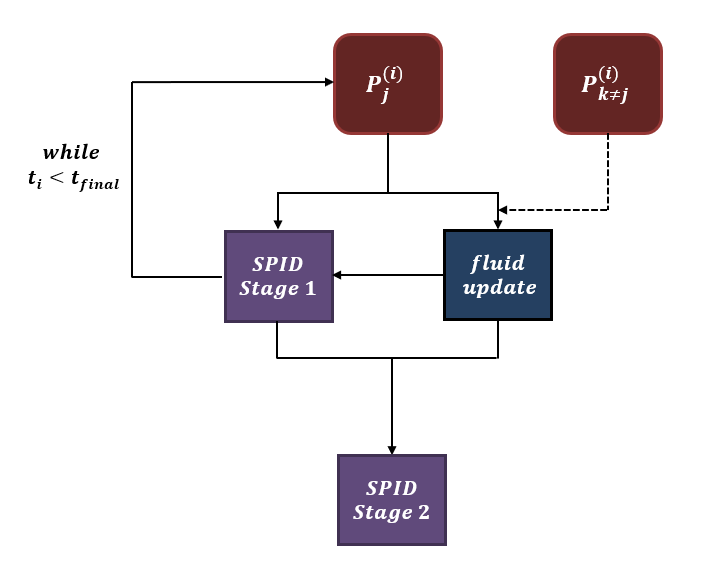}}
   \caption{\label{fig:fluid_modified_flowchart} \heather{CFD analysis with the modified SPID algorithm detailed in Section~\ref{modified_spid}.}}
  \end{figure}

The structure of a CFD solver with this ID implementation is shown in Figure~\ref{fig:fluid_modified_flowchart}. The first stage of the SPID implementation is performed during the fluid solve, with only the second stage taking place after the completion of the fluid solve.

This is detailed in Algorithm~\ref{alg:cfd_w_mod_SPID}. \heather{Similar to} Algorithm~\ref{alg:cfd_w_SPID}, the \textit{update\_fluid\_step} and \textit{save\_subsampled\_data} tasks are launched for each processor at each time step. However, there is now an additional task, \textit{first\_SPID}, which is launched if the solver time is at some critical time $t_{crit}$. \heather{This critical time, which occurs each time a block of time step data has been subsampled and saved, is defined as
  \begin{equation}
  t_{crit} \equiv \Delta t i_{crit},
  \end{equation}
  where $i_{crit}$ is an iteration index that satisfies
  \begin{equation}
  i_{crit} \bmod{\left( \frac{n}{N} \right)} = 0.
  \end{equation}
  }
This implementation reduces both memory overhead and computation time; since the ID algorithm is applied to each processor independently of the others, the computations for the first stage of the SPID can be performed while the flow solver is waiting for the communication of stencil data from neighboring regions for the next time step. Though the \textit{second\_SPID} task is not launched until the completion of the fluid solve, it is operating on a much smaller matrix than that of Algorithm~\ref{alg:cfd_w_SPID}.

\begin{algorithm}[h]
\caption{Task-Parallel CFD Solver with modified SPID}	
\label{alg:cfd_w_mod_SPID}
\begin{algorithmic}[1]
\While{$t < t_{final}$}
\For{$\text{id} = 0, N$}  
  \State $\text{update\_fluid\_step(block[id], stencil[id]})$
  \State $\text{save\_subsampled\_data(block[id])}$
    \If{$t == t_{crit}$}
      \State $\text{first\_SPID(subsampled\_block[id])}$
    \EndIf
\EndFor
\EndWhile
\For{$\text{id} = 0, N$}  
  \State $\text{second\_SPID(subsampled\_block[id])}$
\EndFor
\end{algorithmic}
\end{algorithm}

\section{The Legion programming model} \label{Legion_sect}
\label{sec:legion}
Legion, an open-source, collaborative effort between Stanford University, \alec{Los Alamos National Laboratory (LANL)}, NVIDIA Research, and \alec{Stanford Linear Accelerator Center (SLAC)}, is a parallel programming model that uses task-based parallelism to create highly efficient and portable code for high-performance computing applications \cite{Legion_paper}. The main unit of abstraction in Legion is the \textit{logical region}. Logical regions allow \heather{a} precise definition of how data is being used by the various tasks of an application, and can be further partitioned into various \textit{subregions}. Legion builds a task graph that analyzes the relationship between these subregions and the tasks that access them, and the Legion runtime \textit{dynamically} assigns eligible tasks to available computing units (CPUs, GPUs, etc.). Legion has been specifically designed to target heterogeneous distributed computer architectures with deep memory hierarchies \cite{Legion_paper}. We highlight several areas that demonstrate how these considerations have been taken into account; each provides users targeting these architectures with distinct advantages over traditional parallel programming methodologies.

\subsection{Performance with Legion}
\begin{figure}[!h]
       \center{\includegraphics[width=0.49\textwidth]
       {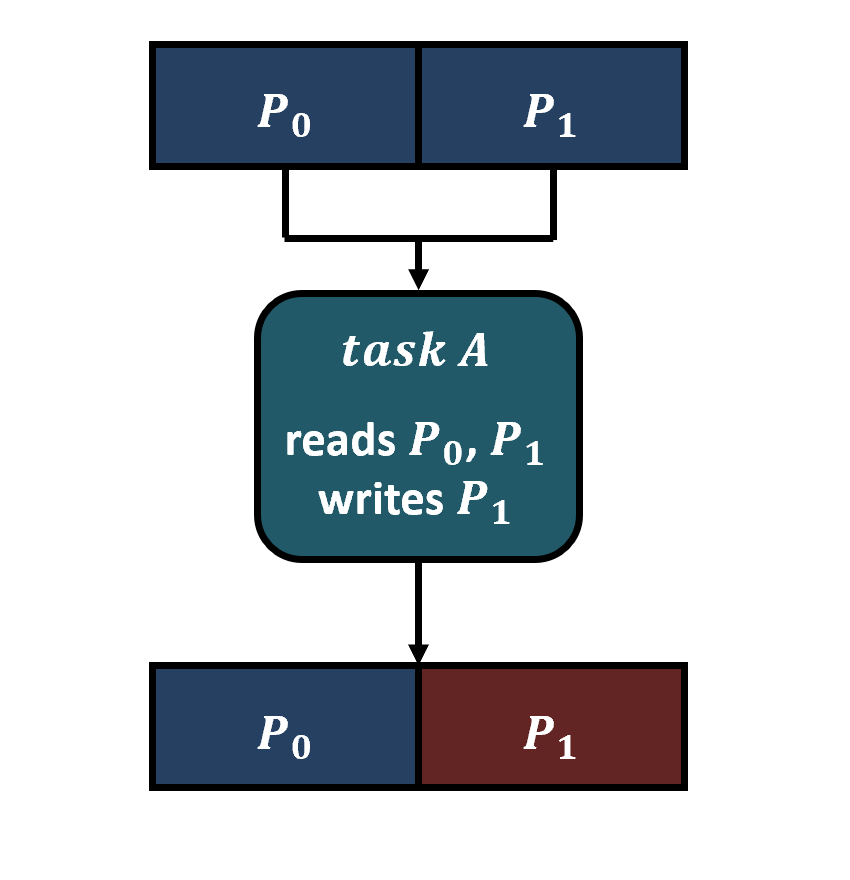}}
       \caption{\label{fig:task_priv} Task privileges in Legion; the data in region $P_0$ is unable to be overwritten, but the data in region $P_1$ may be altered.}
      \end{figure}

Parallelism is implicitly extracted from the task graph that is maintained by the Legion runtime. This task graph is constructed via explicit user declarations of the data a task will require access to, as well as privileges that describe how this data is used (e.g, read, write) within the task\heather{. A} simple example of these declarations is shown in Figure~\ref{fig:task_priv}. The Legion runtime uses this task graph to determine the ordering of the tasks during program execution; its overhead is minimized by performing runtime analysis in parallel with the application \cite{Runtime_paper}. Because the runtime dynamically assigns tasks, it is guaranteed to fully adapt to the inherent unpredictability of machine architectures.
This is in direct contrast to an application that uses MPI+X, where MPI is used for inter-process communication and ``X" refers to a programming model used for finer-grain parallelism (CUDA, OpenMP, Pthreads, and so forth) \cite{PAW-ATM}. In these applications, the task and data placements, as well as synchronization, are the responsibility of the user. For complex applications, these placements can become difficult, and it may be hard to program optimal configurations.

\begin{figure*}[!h]
       \center{\includegraphics[width=1.0\textwidth]
       {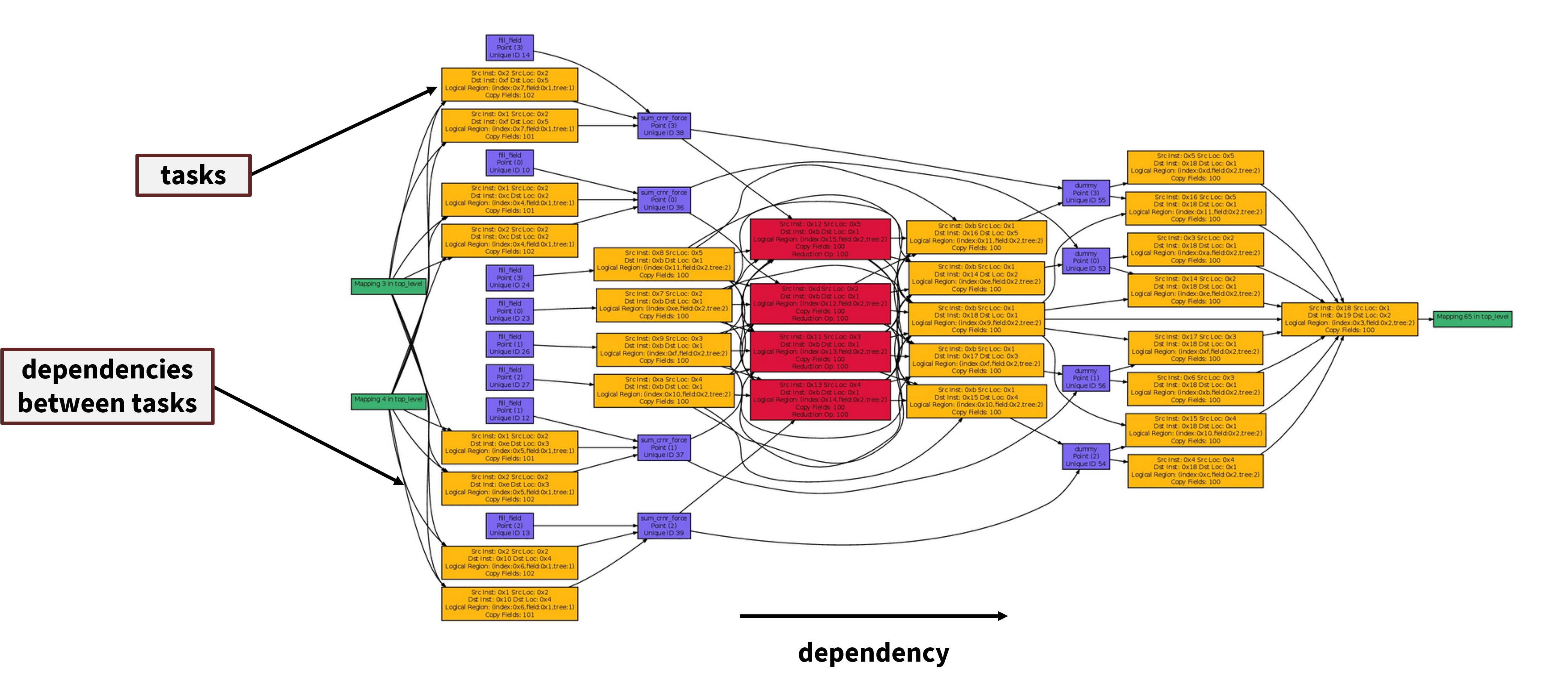}}
       \caption{\label{fig:legion_spy} An example of the LegionSpy debugging tool \cite{spy_ex}; this tool visualizes the task graph of an application. Blocks represent events and arrows represent dependencies between them.} 
      \end{figure*}

There are several tools that allow users to monitor the execution of their Legion application on a machine and assess parallel performance. The first, \heather{LegionSpy} (Figure~\ref{fig:legion_spy}), visualizes events (such as tasks or data copies) and the dependencies between them. This can be used to confirm that the task graph is indeed correct; an unexpected result may indicate that one or more dependencies between tasks or privileges on data may have been declared improperly. 
      
\begin{figure*}[!h]
       \center{\includegraphics[width=1.0\textwidth]
       {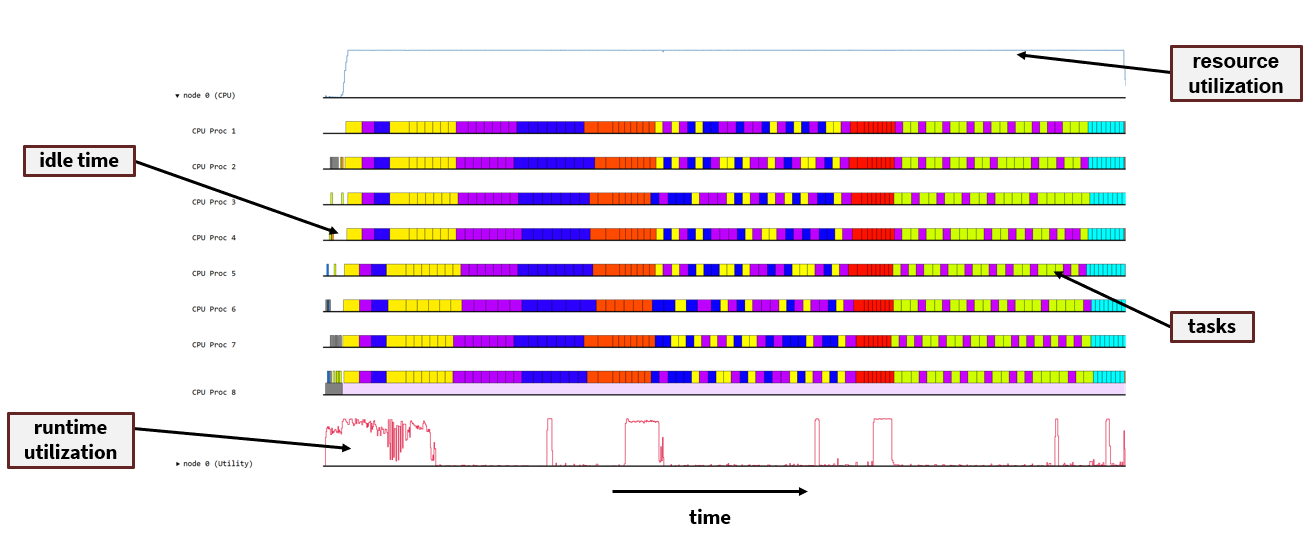}}
       \caption{\label{fig:legion_prof} An example of the LegionProf profiling tool. This tool provides detailed information about the specific execution of an application on a machine, including the task distribution across compute processors, overall resource utilization, and runtime activity.}
      \end{figure*}

The second tool is \heather{LegionProf}, which visualizes the execution of an application across a set of provided compute resources. An example profile is shown in Figure~\ref{fig:legion_prof}. Each colored block represent a completed task, and white space indicates idle time. The tool also provides total resource and runtime utilization. 

\subsection{Portability with Legion}
The user is not responsible for explicitly declaring processors \heather{on which} tasks are run, or where logical regions are placed within memory hierarchies \heather{available to} their application. Instead, a mapping interface is provided that makes machine-specific execution decisions. This mapper is responsible for determining things like
\begin{itemize}
    \item which processor and node a task will run on,
    \item which variant (such as CPU or GPU) of a task will be used,
    \item the exact location in the machine's memory hierarchy all data associated with a task will be laid out.
\end{itemize}

This mapping capability allows a user to have fine-tuned performance control of their application without accessing or modifying their main code; two distinct, valid mapping sets (such as for two different architectures) will produce an identical application solution. Because of this, Legion applications can be easily ported to new architectures, with performance \textit{orthogonal to accuracy}.  Additionally, if a user does not want to develop a custom mapper, there is a default mapper option available. This default mapper makes generic decisions and allows applications to be developed in a time-efficient manner. 

Portability is greatly increased, as the user is only required to write a single Legion application. This application can be run on CPUs, GPUs, a mix of CPUs and GPUs, shared memory systems, and distributed memory systems without modification. This is unlike MPI+X applications, which must be modified and/or rewritten for each individual type of processor and architecture. Because these features are directly embedded within the application, there is also the possibility that solution inaccuracies are introduced to the application each time it is modified. This is not an issue with Legion applications, as the mapper is a standalone component. 

\subsection{Simplicity with Legion} 
Since parallelism is implicitly extracted, Legion applications are written in sequential, easy-to-read code. This also means that, in general, a Legion application requires fewer lines of code than an application that uses MPI+X. These savings can be substantial; in large multi-physics codes, up to 20\% of the total code can be dedicated to explicit communication or synchronization instructions. Legion is a C++ library; users also have the option of writing their applications in Regent, which is a high-level programming language built on top of Legion \cite{Regent_paper}. 

Regent contains several task annotations that allow the user to easily control how the application is run on a machine, even if they are using the default mapping feature. For example, users can generate a GPU-enabled variant of a task by simply adding \_\_demand(\_\_cuda) to the top of a task. Annotations can also be applied to for-loops; users can create OpenMP variants, or generate vectorization. These features, as well as the high-level, sequential nature of Regent itself, allow for easier understanding and troubleshooting of existing applications than traditional MPI+X applications.

\section{Numerical Experiments}
\label{sec:numerical_experiments}
The test cases below demonstrate the successful application of ID for data compression into two canonical Taylor-Green vortex problems. These examples were selected because their solution is provably low rank; application of ID to these data matrices should result in ideal compression values. Further, the first problem presented (a Taylor-Green vortex) has an analytic solution. \alec{Therefore, failure to compress data generated from simulations of these systems would be due to the algorithmic implementation itself.}  

\subsection{Analytical Taylor-Green Vortex}

The two-dimensional incompressible Taylor-Green vortex, a decaying vortex with unsteady flow, is an exact closed form solution of the incompressible Navier-Stokes equations in a Cartesian coordinate system.

Let $\textbf{u}=(u_1, u_2)$ be the velocity components of the flow, $p$ the pressure field of the flow, $\rho$ the density, and $\nu$ the kinematic viscosity of the fluid. The exact Taylor-Green solution of this system on the doubly periodic domain $0 \leq x_1,x_2 < 2\pi$ is
\begin{align}
    u_1(x_1, x_2, t) &= \sin{(x_1)} \cos{(x_2)} \exp{(-2 \nu t)}, \\
    u_2(x_1, x_2, t) &= -\cos{(x_1)} \sin{(x_2)} \exp{(-2 \nu t)}, \\
    p(x_1, x_2, t) &= \frac{\rho}{4} \left(\cos{(2 x_1)} + \sin{(2 x_2)} \right) \exp{(-4 \nu t)} .
\end{align}

\begin{figure}[!h]
       \center{\includegraphics[width=0.49\textwidth]
       {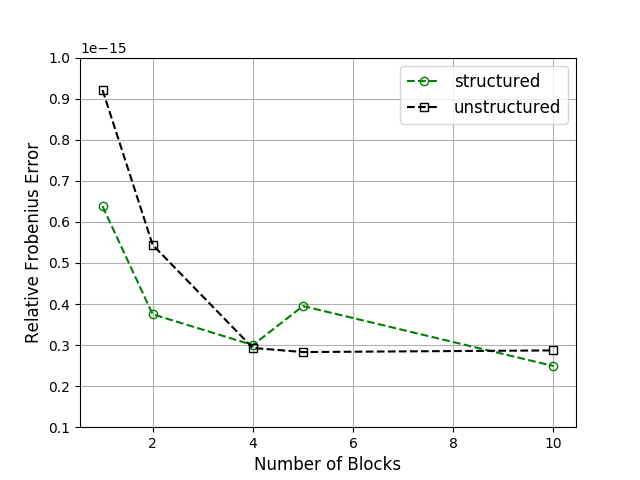}}
       \caption{\label{fig:tg_1} Error in the $u_1$ component of velocity as a function of blocks in the $x_1$-direction.}
      \end{figure}

\begin{figure}[!h]
       \center{\includegraphics[width=0.49\textwidth]
       {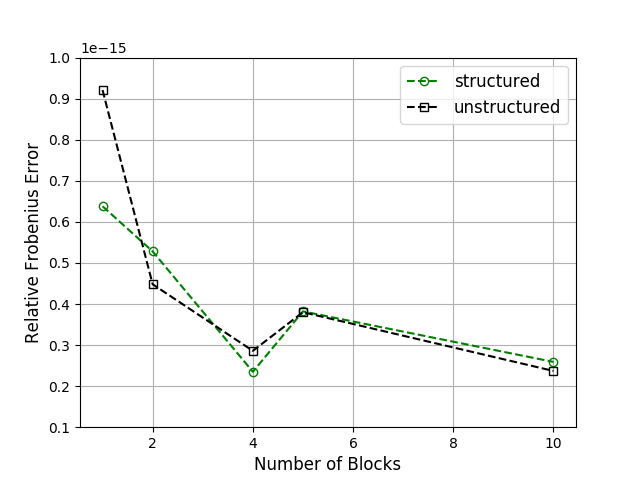}}
       \caption{\label{fig:tg_2} Error in the $u_1$ component of velocity as a function of blocks in the $x_2$-direction.}
      \end{figure}

\begin{figure}[!h]
       \center{\includegraphics[width=0.49\textwidth]
       {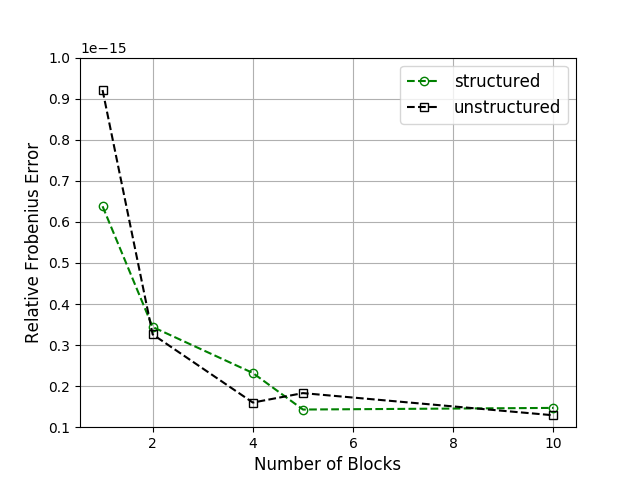}}
       \caption{\label{fig:tg_3} Error in the $u_1$ component of velocity as a function of blocks in the $x_1$ and $x_2$-direction.}
      \end{figure}

The ID algorithm described in Section~\ref{id_imp} was applied to a Legion implementation of the Taylor-Green solution \alec{defined on both a structured and an unstructured, randomly generated grid}. The reconstructed (compressed) solution $\bm{A}_{approx}$ was then compared to the exact solution $\bm{A}_{exact}$ at the grid points, and the normalized Frobenius norm (matrix entry-wise 2-norm) of the error
\begin{equation}
    \text{Relative Frobenius Error} = \frac{\norm{\bm{A}_{\text{exact}} - \bm{A}_{\text{approx}}}_{F}}{\norm{\bm{A}_{\text{exact}}}_{F}},
\end{equation}
was computed. Error results for the data compression of the $u_1$ component of velocity as a function of domain partitions for $20^2$ domain points \alec{for both the structured and unstructured grids} are shown in Figures~\ref{fig:tg_1}~-~\ref{fig:tg_3}. For a total simulated time of 10 s (100 total ``time steps" of the analytical solution), the resulting rank of the ID approximation is one (which corresponds to a compression factor $\approx80$). For both grids, the normalized Frobenius norm of the error is on the order of machine precision. Similar accuracy and compression factor results are observed for the $u_2$ component of velocity, as well as the pressure. 

This demonstrates that the SPID algorithm can achieve a large compression ratio even when there is no \alec{``sorting"} to the underlying points of the spatial domain. It is also important to note that the accuracy of the approximation increased as the number of spatial partitions increased; partitioning the spatial domain can both increase the computational efficiency and compression ratio without sacrificing accuracy.

\begin{figure*}[!htb]
       \centering
       \begin{tabular}{@{}ccc@{}}
       \subfloat[Exact solution, structured grid.]
       {\label{fig:ex_struct} \includegraphics[width=0.49\textwidth]{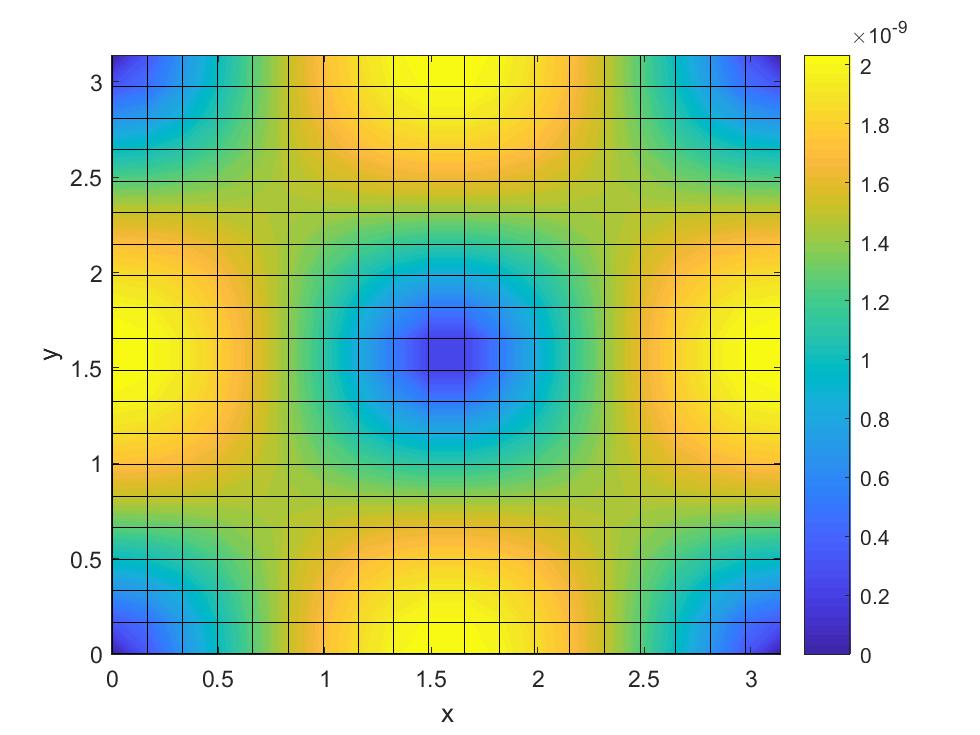}}
      
       \subfloat[Reconstructed solution, structured grid.]
       {\label{fig:app_struct} \includegraphics[width=0.49\textwidth]{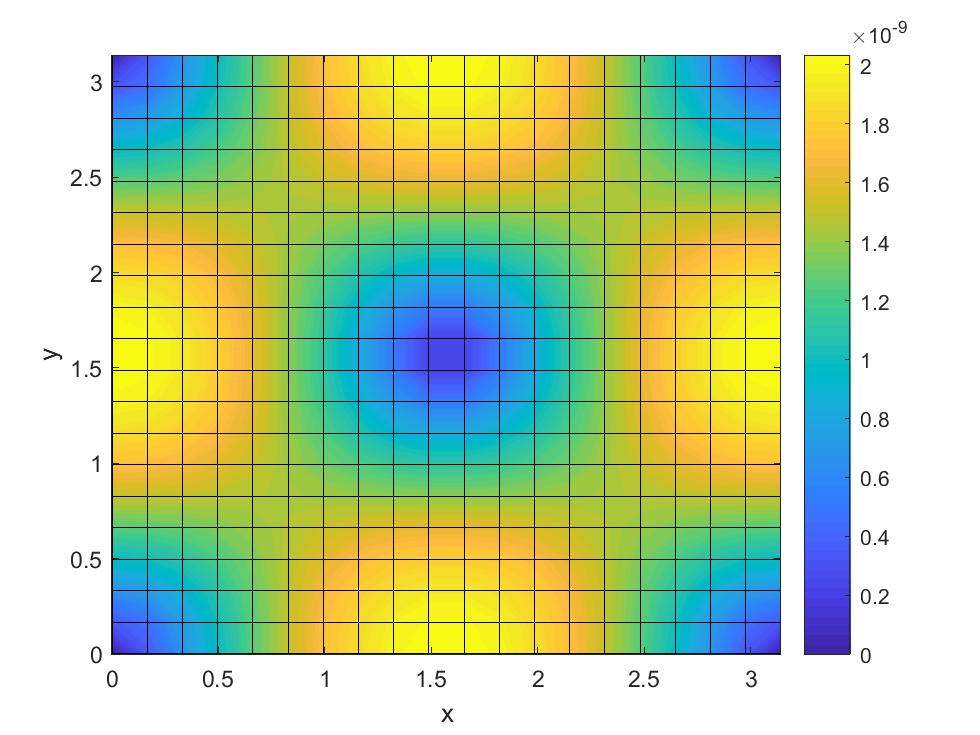}} \\

       \subfloat[Exact solution, unstructured grid.]
       {\label{fig:ex_unstruct} \includegraphics[width=0.49\textwidth]{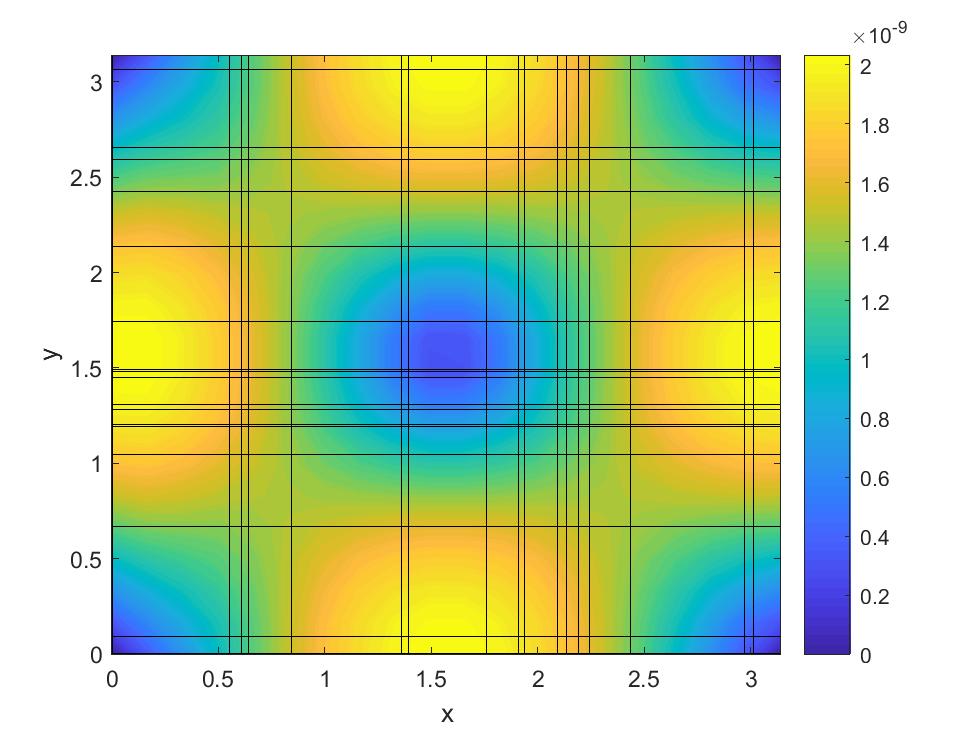}}
      
       \subfloat[Reconstructed solution, unstructured grid.]
       {\label{fig:app_unstruct} \includegraphics[width=0.49\textwidth]{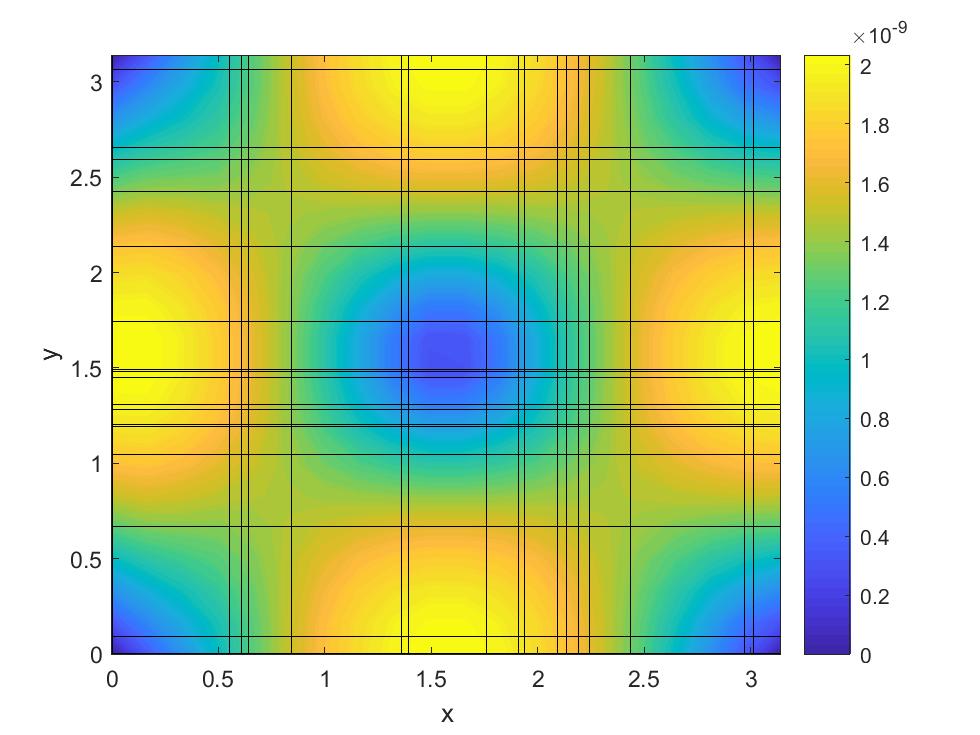}}
\end{tabular}
       \caption{\label{fig:grid_solutions} Exact and reconstructed solutions for the velocity magnitude of a $20^2$ domain on a structured grid (\protect\subref{fig:ex_struct} and \protect\subref{fig:app_struct}) and an unstructured grid (\protect\subref{fig:ex_unstruct} and \protect\subref{fig:app_unstruct}); total simulated time is 10 s.}
      \end{figure*}

The exact and reconstructed velocity magnitude at $t=10$ s for \heather{both the $20^2$ structured and unstructured grid domains} with 10 partitions in both the $x_1$- and $x_2$-directions can be seen in Figure~\ref{fig:grid_solutions}. The normalized Frobenius norm of the error for this velocity magnitude is $\mathcal{O}(10^{-16})$. Even though there is a large number of partitions in the domain, each with their own individual ID approximation, there is no ``disjointedness" apparent in this reconstructed solution.

\subsection{High-order Navier-Stokes solver} \label{NSID}
Our second test case employs a high-order Cartesian solver for the Navier-Stokes equations that is used as an off-body solver for rotorcraft wake simulation \cite{cart_p_paper_2}, as well as a test bed for the development of novel numerical methods \cite{cart_p_paper}. It solves the conservation form of the \alec{3D} Navier-Stokes equations
\begin{equation}
\frac{\partial q}{\partial t} +  \frac{\partial f_1}{\partial x_1} + \frac{\partial f_2}{\partial x_2} + \frac{\partial f_3}{\partial x_3}  = \frac{\partial g_1}{\partial x_1} + \frac{\partial g_2}{\partial x_2} + \frac{\partial g_3}{\partial x_3} ,
\end{equation}
where $q$ is the vector of conserved variables, $f_i$ is the convective flux vector in the $i^{th}$ direction, and $g_i$, is the viscous flux vector in the $i^{th}$ direction:
  \begin{equation}
    q = \begin{bmatrix}
          \rho \\
          \rho u_1 \\
          \rho u_2 \\
          \rho u_3 \\
          \rho e
         \end{bmatrix} , \quad
    f_i = \begin{bmatrix}
          \rho (u_i-\dot{x}_i) \\
          \rho u_1(u_i-\dot{x}_i) + p \\
          \rho u_2(u_i-\dot{x}_i) \\
          \rho u_3(u_i-\dot{x}_i) \\
          \rho e(u_i-\dot{x}_i)+pu_i
         \end{bmatrix} , \quad
 g_i = \begin{bmatrix}
          0 \\
          \tau_{i1} \\
          \tau_{i2} \\
          \tau_{i3} \\
          u_j \tau_{ij}-\textbf{k}_i
         \end{bmatrix};
  \end{equation}
  
  \noindent $\textbf{u} = (u_1, u_2, u_3)$ is the fluid velocity, $\dot{\textbf{x}} = (\dot{x}_1, \dot{x}_2, \dot{x}_3)$ is the reference frame velocity, $\tau$ is the viscous stress tensor, $\textbf{k}$ is the heat flux, $\rho$ and $p$ are the fluid density and pressure, respectively, and $e$ is the total energy. \\
  
\noindent A Newtonian fluid assumption is used, so the terms of the viscous stress tensor are defined as
\begin{equation}
\tau_{ij} = \mu \left(\frac{\partial \textbf{u}_i}{\partial x_j}+\frac{\partial \textbf{u}_j}{\partial x_i} \right) - \frac{2}{3} \mu \delta_{ij} \frac{\partial \textbf{u}_k}{\partial x_k} .
\end{equation}

\noindent The dynamic viscosity $\mu$ is considered to be a function of temperature according to Sutherland's Law
\begin{equation}
    \mu (T) = \mu_{ref} \left( \frac{T}{T_{ref}} \right)^{\frac{3}{2}} \frac{T_{ref} + S}{T + S} ,
\end{equation}
where $S$ is the Sutherland temperature, $T_{ref}$ is a reference temperature, and $\mu_{ref}$ is the dynamic viscosity at the reference temperature. \\

\noindent The entries of the heat flux vector are
\begin{equation}
    \textbf{k}_i = -\kappa \frac{\partial T}{\partial x_i} ,
\end{equation}
with the thermal conductivity $\kappa$ found from the definition of the Prandtl number $Pr$
\begin{equation}
\alec{\kappa} = \mu \frac{c_p}{Pr} ,
\end{equation}
where $c_p$ is the constant pressure specific heat. The system is closed with the equation of state for an ideal gas
\begin{equation}
p = (\gamma - 1) \left(\rho e -\frac{1}{2} \rho (u_1^2+u_2^2+u_3^2) \right) ,
\end{equation}
and the speed of sound is defined as
\begin{equation}
a = \sqrt{\gamma \frac{p}{\rho}} ,
\end{equation}
where $\gamma \equiv \frac{c_p}{c_v}$ is the ratio of the constant pressure and constant volume specific heats.

\subsubsection{Spatial Discretization}

The Navier-Stokes equations are transformed into ordinary differential equations by discretizing in space on the Cartesian grid with cell spacing $(\Delta x_1, \Delta x_2, \Delta x_3)$, so that
\begin{equation} \label{eq:NS}
\frac{d q_j}{dt} = \sum_{i=1}^{3} \left( \frac{\partial g_i}{\partial x_i} - \frac{\partial f_i}{\partial x_i} \right) \biggr\rvert _{j} .
\end{equation}

       \begin{figure}[!h]
       \center{\includegraphics[width=0.5\textwidth]
       {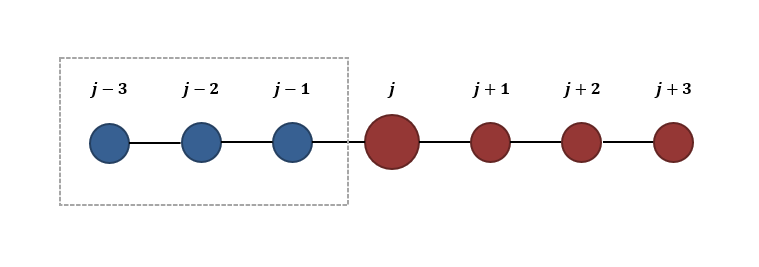}}
       \caption{\label{fig:stencil} Seven-point stencil for one direction of the spatial domain; this requires three fringe points that must be exchanged between boundaries for each spatial direction.}
      \end{figure}

The solver is \alec{based on} sixth-order accurate inviscid flux discretizations and fourth-order accurate viscous flux discretizations. There is also a fifth-order accurate artificial dissipative flux term introduced into the inviscid flux reconstructions for stability purposes. These terms all use a seven-point stencil in the spatial domain (Figure~\ref{fig:stencil}); for example, the inviscid flux gradients at the cell centers are approximated as
\begin{equation} \label{eq:inviscid}
\frac{\partial f_i}{\partial x_i} \biggr\rvert _{j} \approx \frac{ \hat{f}_{i_{j+\frac{1}{2}}} -  \hat{f}_{i_{j-\frac{1}{2}}}}{\Delta x_i} ,
\end{equation}

\noindent where $\hat{f}_{i_{j+\frac{1}{2}}}$ and $\hat{f}_{i_{j-\frac{1}{2}}}$ are the fluxes at the cell faces in the $j^{th}$ direction. The fluxes at the cell faces are defined as

\begin{equation} 
\hat{f}_{i_{j+\frac{1}{2}}} = \frac{1}{60}(f_{i_{j-2}}-8 f_{i_{j-1}} +37 f_{i_{j}} +  37 f_{i_{j+1}} - 8 f_{i_{j+2}} + f_{i_{j+3}}) .
\end{equation}
        
These flux reconstructions result in three fringe points at each boundary in the partitioned numerical domain.        
        
\subsubsection{Temporal Discretization}
        
The Navier-Stokes equations are explicitly advanced in time using a low-storage $3^{rd}$-order Runge-Kutta method \cite{rk3}. For the following ODE:

\begin{equation}
\frac{d q}{dt} = -\textbf{R}(q) ,
\end{equation}
the solution at the next time step, $q^{(n+1)}$, is found from
\begin{align}
q^*&=q^{(n)}-a_1 \Delta t \textbf{R}(q^{(n)}) ,\\
q^{**}&=q^{(n)}-a_2 \Delta t \textbf{R}(q^{(n)}) ,\\
q^{***}&=q^{*}-a_3 \Delta t \textbf{R}(q^{**}) ,\\
q^{(n+1)}&=q^{*}-a_4 \Delta t\textbf{R}(q^{***}) ,
\end{align}
where $a_1=\frac{1}{4}$, $a_2=\frac{8}{15}$, $a_3=\frac{5}{12}$, and $a_4=\frac{3}{4}$. \\

Overall, there is a high amount of communication required for this solver; this three-stage process for the time-advancement scheme amplifies the already large amount of data that must be communicated between neighboring partitions due to the seven-point stencil in the spatial domain.

\subsubsection{Legion Implementation and ID Extension}
This solver was re-engineered in the Legion framework, and data compression via the SPID algorithm was added. To test how well the Legion implementation performs, a compressible Taylor-Green vortex with the following initial conditions 
\begin{align}
u(t_0) &= V_0 \sin{\bigg(\heather{\frac{x_1}{L}}\bigg)}\cos{\bigg(\heather{\frac{x_2}{L}}\bigg)}\cos{\bigg(\heather{\frac{x_3}{L}}\bigg)} ,\\
v(t_0) &= -V_0\cos{\bigg(\heather{\frac{x_1}{L}}\bigg)}\sin{\bigg(\heather{\frac{x_2}{L}}\bigg)}\cos{\bigg(\heather{\frac{x_3}{L}}\bigg)} ,\\
w(t_0) &= 0  ,\\
p(t_0) &= p_0 +\frac{\rho_0 V_0^2}{16} \bigg( \cos{\bigg(2\heather{\frac{x_1}{L}}\bigg)}+\cos{\bigg(2\heather{\frac{x_2}{L}}\bigg)}\bigg)\bigg(\cos{\bigg(2\heather{\frac{x_3}{L}}\bigg)}+2  \bigg) ,
\end{align}
was simulated on the triply-periodic domain $0 \leq x_1,x_2,x_3 < 2 \pi L$ with $L=1$, $V_0=1$, $\rho_0=1$, $p_0=100$, and  $T_0=300 \text{K}$. The viscosity was adjusted so that the Reynolds number is $\mathcal{O}(100)$ to \alec{maintain laminar flow.}
 
\begin{figure}[!h]
       \center{\includegraphics[width=0.49\textwidth]
       {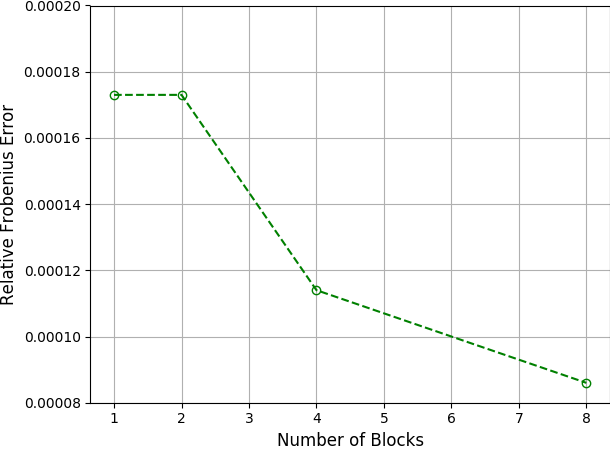}}
       \caption{\label{fig:comp_TG_error} Error in the $u_1$ component of velocity as a function of blocks in the $x_1$, $x_2$, and $x_3$-directions for 100 time steps of a $64^3$ point domain (0.0001 s of simulated time).}
      \end{figure}

 The accuracy for the $u_1$-component of velocity is shown in Figure~\ref{fig:comp_TG_error} for a $64^3$ point domain and 100 time steps (0.0001 s of simulated time). We observe relative Frobenius norm errors lower than $10^{-3}$ using 1, 2, 4, and 8 blocks to partition the domain in each spatial direction. As the number of blocks is increased, the error decreases  by roughly 50 percent. For each number of partitions, the resulting compressed matrix approximation was rank one, with a compression factor of approximately 100. Similar accuracy results were seen for other flow parameters.

      \heather{
      \begin{remark}
      To reduce computational expense when repeatedly running large problem sizes, all results presented in this section are for 100 time steps, or 0.0001 s of simulated time. A study of runs simulating much larger amounts of time resulted in identical levels of accuracy and rank of the ID approximation.
      \end{remark}}
     
\begin{figure}[!h]
       \center{\includegraphics[width=0.49\textwidth]
       {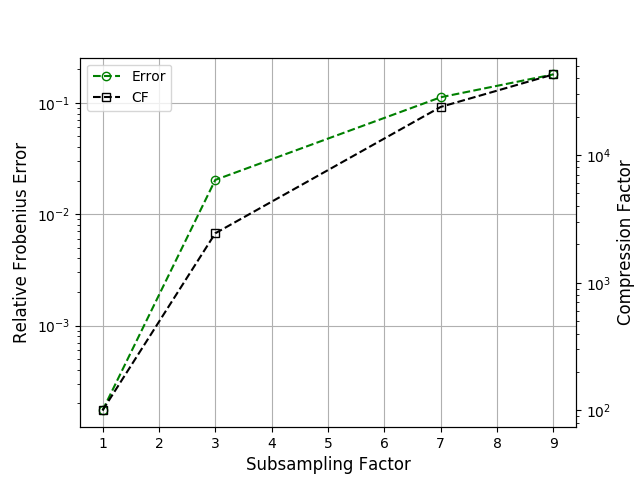}}
       \caption{\label{fig:CF_error_64_3_compare} Error and compression factor in the $u_1$ component of velocity as a function of subsampling in the $x_1$, $x_2$, and $x_3$-directions for 100 time steps of a $64^3$ point domain (0.0001 s of simulated time).}
      \end{figure}

  This accuracy was also measured as a function of subsampling the physical domain. Figure~\ref{fig:CF_error_64_3_compare} shows the error for the $u_1$-component of velocity for the same $64^3$ point domain and 100 time steps, this time with a single domain partition but subsampling intervals of 1, 3, 7, and 9 in each spatial direction. Subsampling increased the overall error in the reconstructed solution, but greatly increased the overall compression ratio. A subsampling interval of 3, while only 2\% different from the full numerical solution, had a compression ratio of approximately 2400, compared to approximately 100 when no subsampling is used.

\begin{figure*}[!ht]
       \centering
       \begin{tabular}{@{}ccc@{}}
       \subfloat[Full numerical solution.]
       {\label{fig:ex_vel_a} \includegraphics[width=0.49\textwidth]{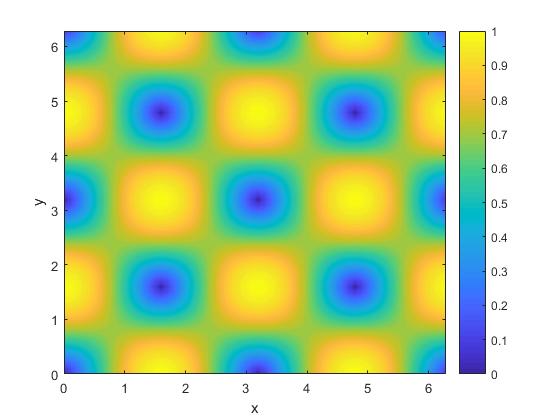}}
      
       \subfloat[Reconstructed solution, no subsampling.]
       {\label{fig:id_vel_b} \includegraphics[width=0.49\textwidth]{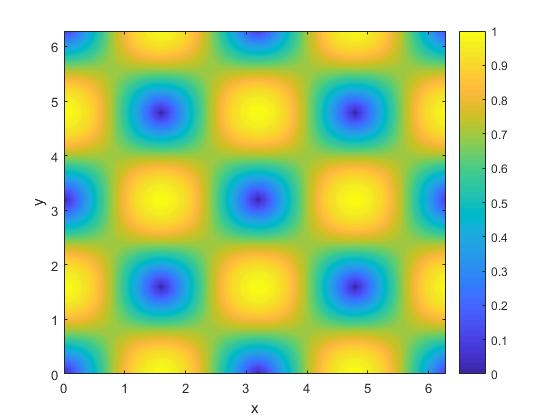}} \\

       \subfloat[Reconstructed solution, subsampling factor = 3.]
       {\label{fig:id_vel_c} \includegraphics[width=0.49\textwidth]{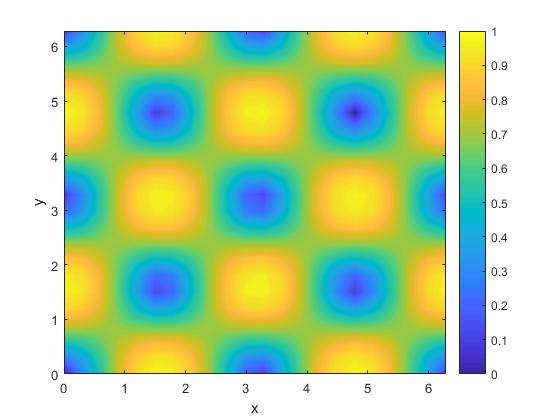}}
      
       \subfloat[Reconstructed solution, subsampling factor = 9.]
       {\label{fig:id_vel_d} \includegraphics[width=0.49\textwidth]{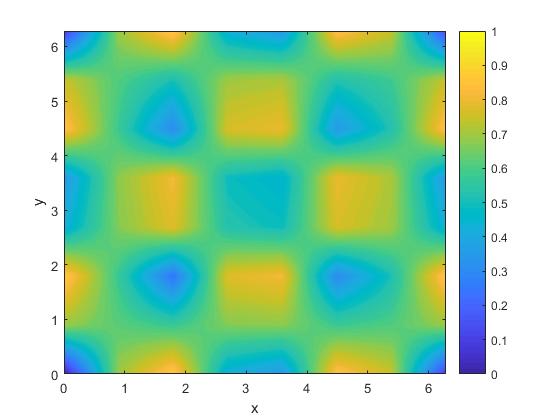}}
\end{tabular}
       \caption{\label{fig:comp_tg_results} Full numerical solution \protect\subref{fig:ex_vel_a} and reconstructed solutions \protect\subref{fig:id_vel_b}, \protect\subref{fig:id_vel_c}, \protect\subref{fig:id_vel_d} for the velocity magnitude of a $64^3$ domain on a structured grid with 1 spatial partition as a function of physical subsampling interval; total simulated time is 0.0001s.}
      \end{figure*}

 Figure~\ref{fig:comp_tg_results} shows the velocity magnitude for the full domain and each of the subsampling intervals. As the subsampling interval increases, the reconstructed solution becomes less accurate, though the overall behavior in the domain is preserved. It is important to note that the accuracy of this result, and SubID results in general, may be affected by the interpolation operator used to transform the subsampled grid to the full grid. In this case, we used trilinear interpolation; another method, such as tricubic interpolation, may yield improved results.

Though the compression and accuracy were ideal, we were also interested in the suitability of the Legion system for a flow solver using ID for data compression. We will assess this on the basis of the three Legion features described in Section~\ref{Legion_sect}.

\begin{figure*}[hp]
       \centering
       \subfloat[NS solver task graph.] 
       {\label{fig:task_graphs_a} \includegraphics[width=1.0\textwidth]{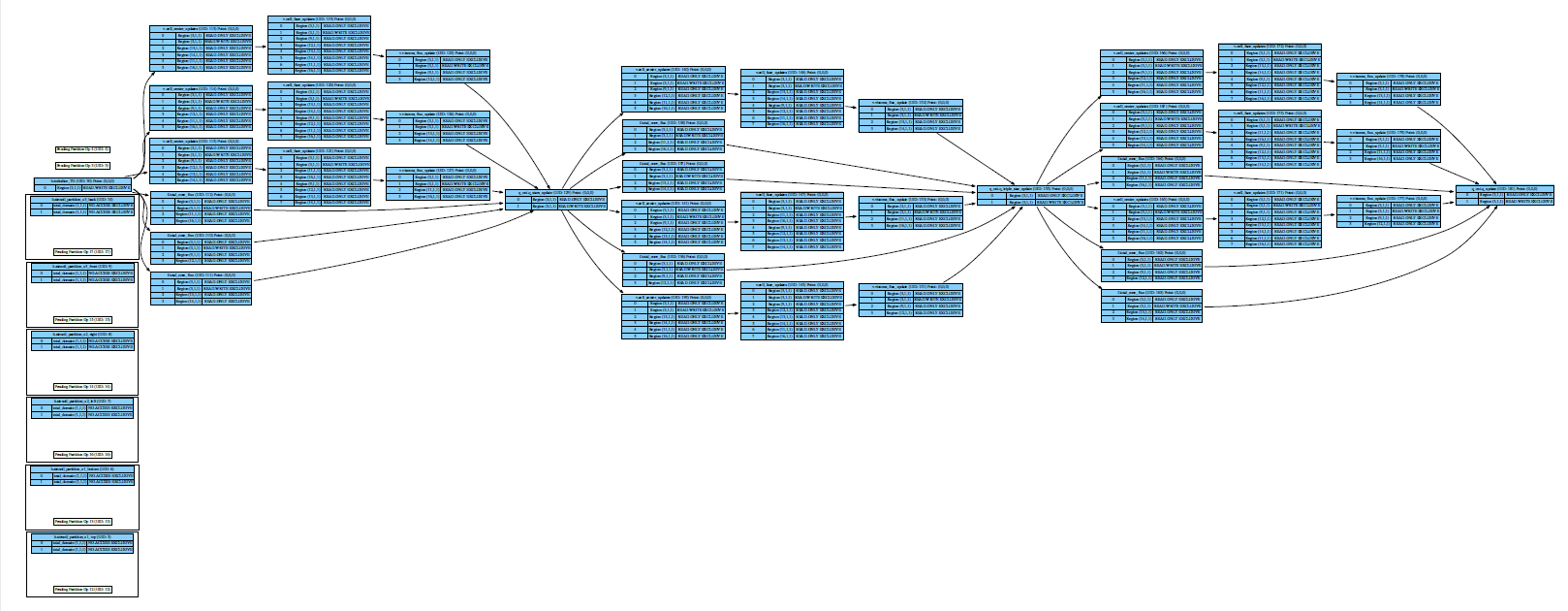}}
      
       \subfloat[NSID solver task graph.] 
       {\label{fig:task_graphs_b} \includegraphics[width=1.0\textwidth]{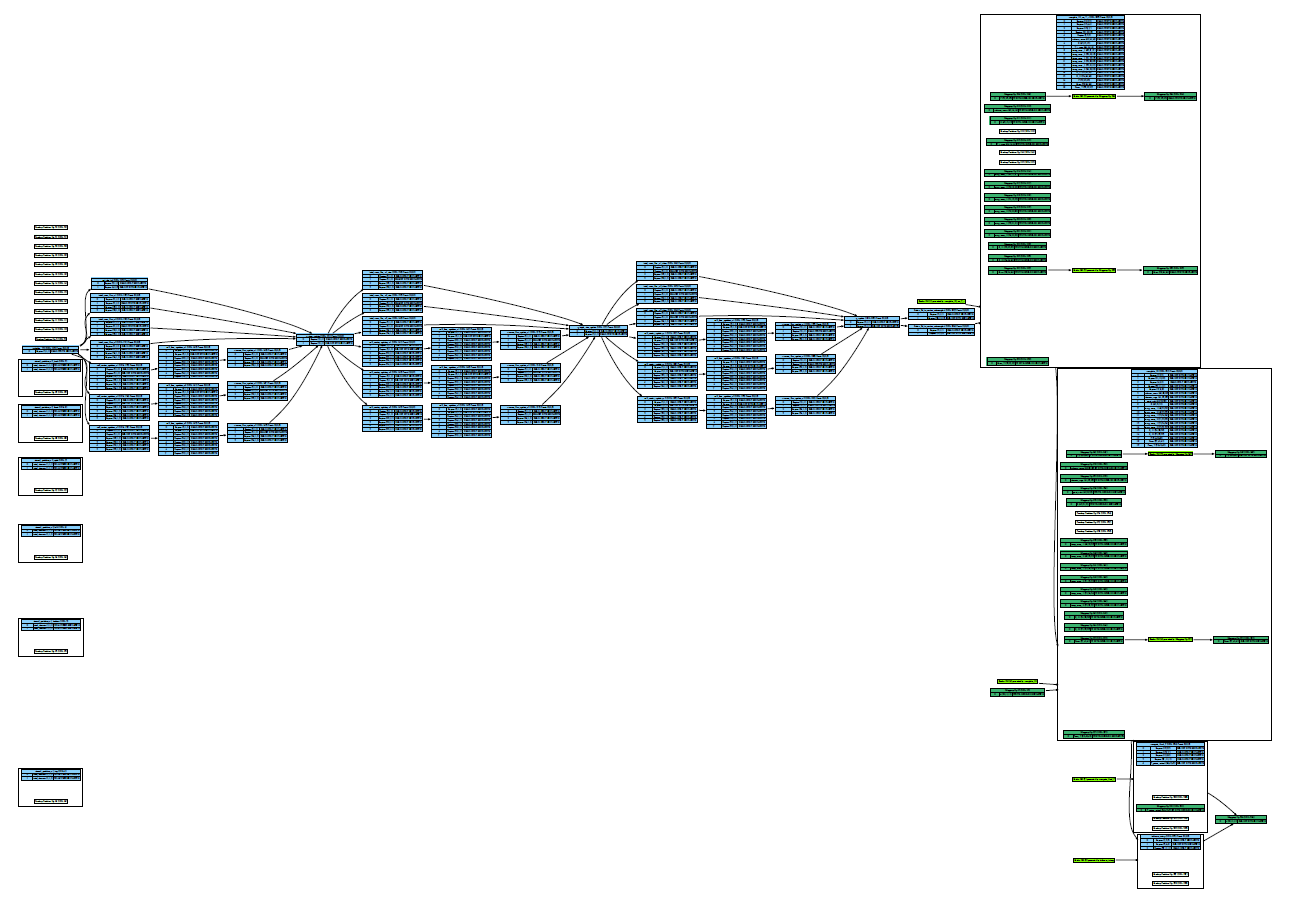}}

       \caption{ \label{fig:task_graphs} LegionSpy task graphs for a single time step of \protect\subref{fig:task_graphs_a} the Navier-Stokes (NS) solver, and \protect\subref{fig:task_graphs_b} the Navier-Stokes Interpolative Decomposition (NSID) solver; blocks represent events and arrows between blocks represent dependencies between events. \heather{There are two possible task executions at the end of the NSID solver time step; if $t \neq t_{crit}$, then the data is simply subsampled and flattened. However, if $t = t_{crit}$, a Stage 1 ID analysis is performed.}}
      \end{figure*}

 \textit{Simplicity:} The Regent programming language was used, and OpenMP and/or CUDA task variants were created for both the flow solver and the SPID algorithm. Figure~\ref{fig:task_graphs} shows LegionSpy task graph analyses for both one time step of the Navier-Stokes (NS) solver and one time step of the Navier-Stokes Interpolative Decomposition (NSID) solver. Each block represents an event required for the time step, and each arrow corresponds to a dependency that requires data communication from the stencil points. As expected, parallelism was implicitly extracted for both codes, eliminating the need for the user to hard-code all communication stages between the flow solver and the SPID stages.
      
\begin{figure}[h]
       \centering
       \subfloat[Strong scaling results for 10 time steps of a 128\textsuperscript{3} point domain with 4\textsuperscript{3} spatial partitions.]
       {\label{fig:scaling_a} \includegraphics[width=0.49\textwidth]{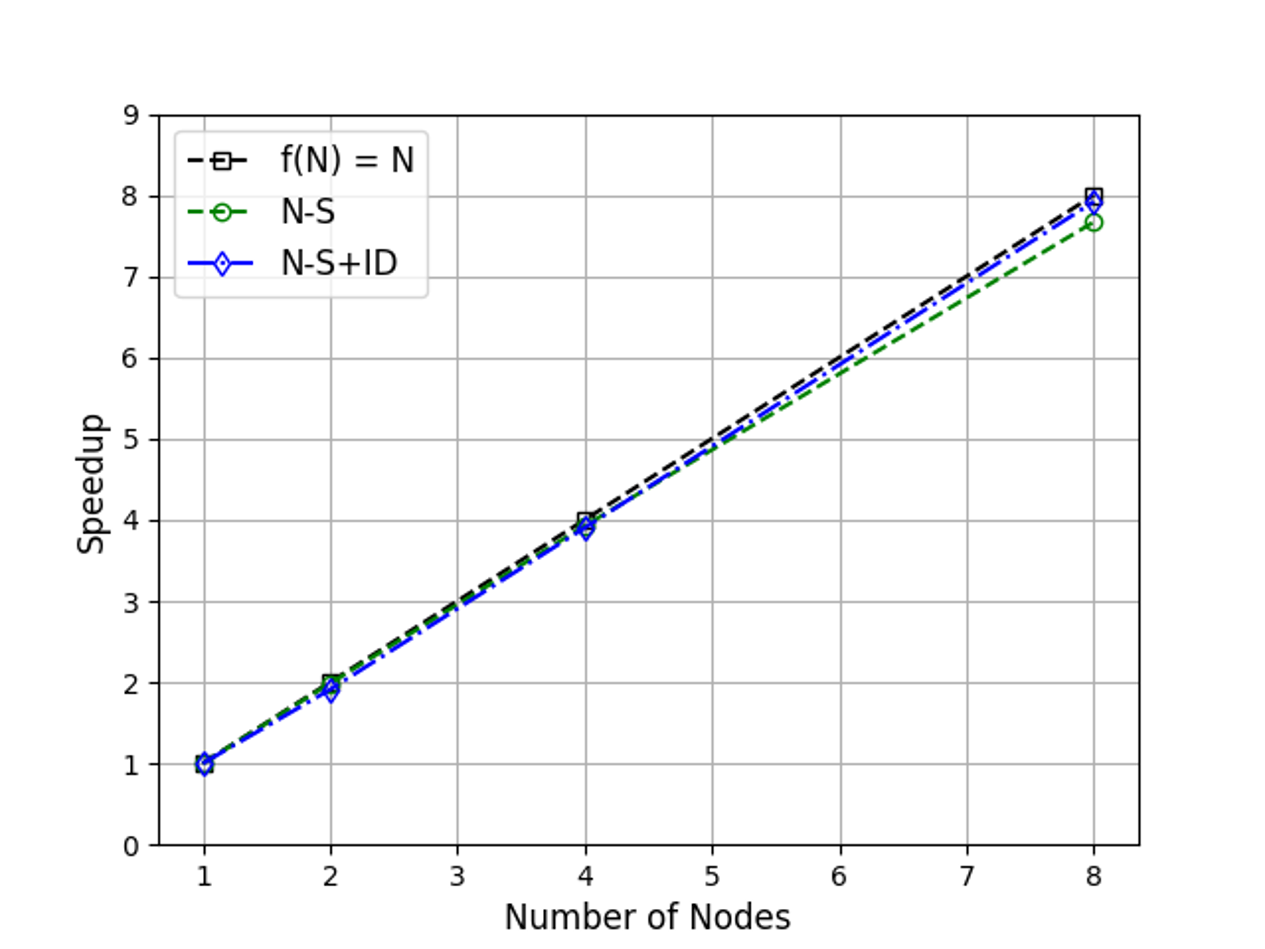}}
      
       \subfloat[The same scaling results, normalized to the single-node runtime of the NS solver.]
       {\label{fig:scaling_b} \includegraphics[width=0.49\textwidth]{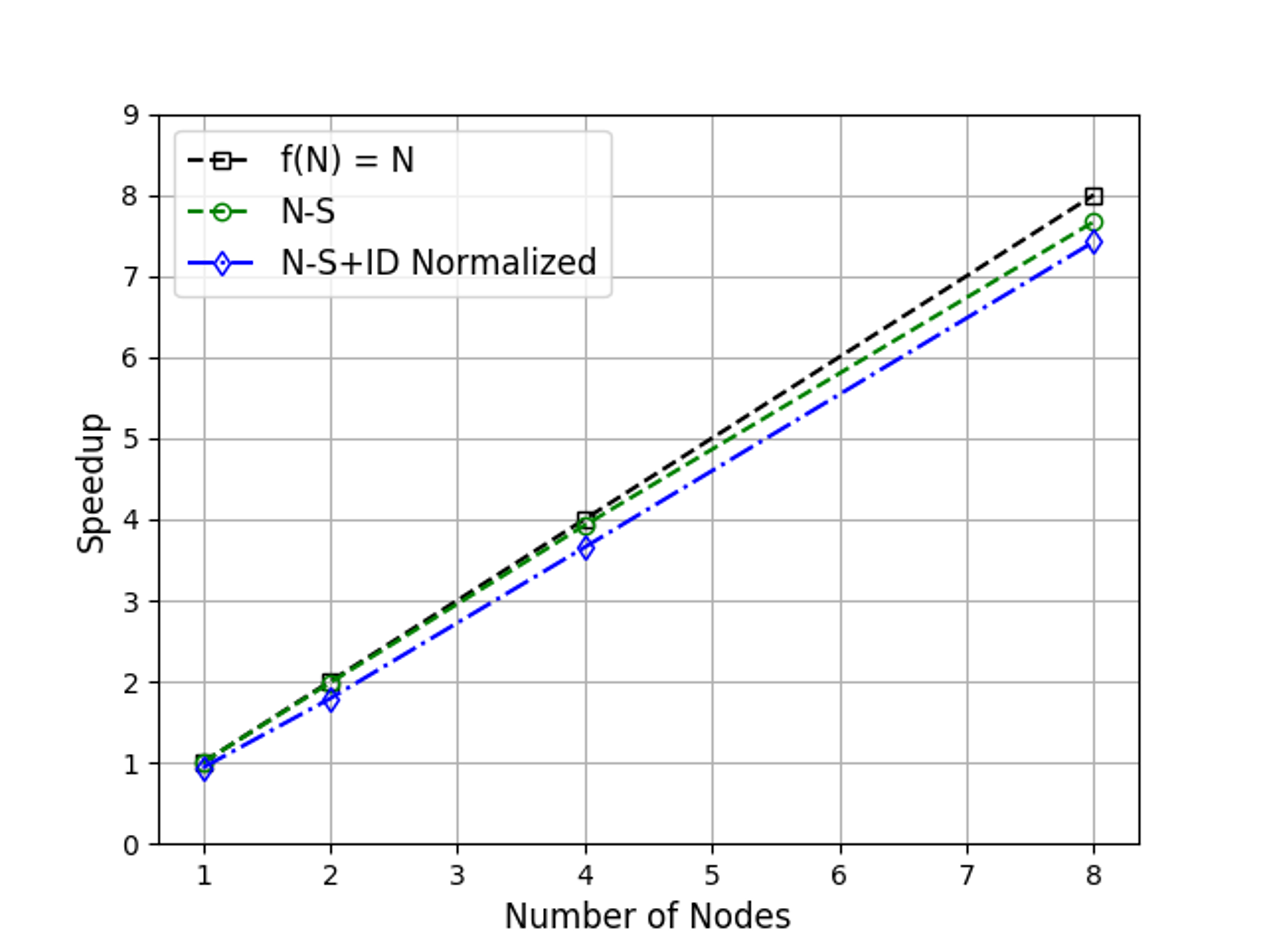}}

       \caption{\label{fig:strong_scaling} Strong scaling results for \protect\subref{fig:scaling_a} NS and \protect\subref{fig:scaling_b} NSID solvers.}
      \end{figure}

\textit{Performance:} From a performance perspective, data compression using ID can only be considered beneficial if it does not negatively impact the scaling of the original solver, and does not significantly increase the overall runtime. Both the NS and NSID solvers were mapped to CPUs to test the strong scaling of the solver for a larger number of nodes. Figure~\ref{fig:scaling_a} shows results for 10 time steps (0.0001 s of simulation time) of a 128\textsuperscript{3} point domain with 4\textsuperscript{3} spatial partitions. Both codes scaled well up to 8 nodes, but the NSID solver scaled slightly better than the Navier-Stokes solver on its own. This is expected, as the SPID algorithm introduces additional work into the NS solver without requiring any additional communication. 

Figure~\ref{fig:scaling_b} shows the scaling normalized to the runtime of the single node Navier-Stokes simulation. The increase in runtime due to the addition of the SPID algorithm was, at most, 10\% that of the original Navier-Stokes code. Thus, introducing the SPID algorithm into the CFD solver did not have negative impacts on scaling or overall runtime.

\begin{figure*}[!h]
       \centering
       \subfloat[]
       {\label{spy_a} \includegraphics[width=1.0\textwidth]{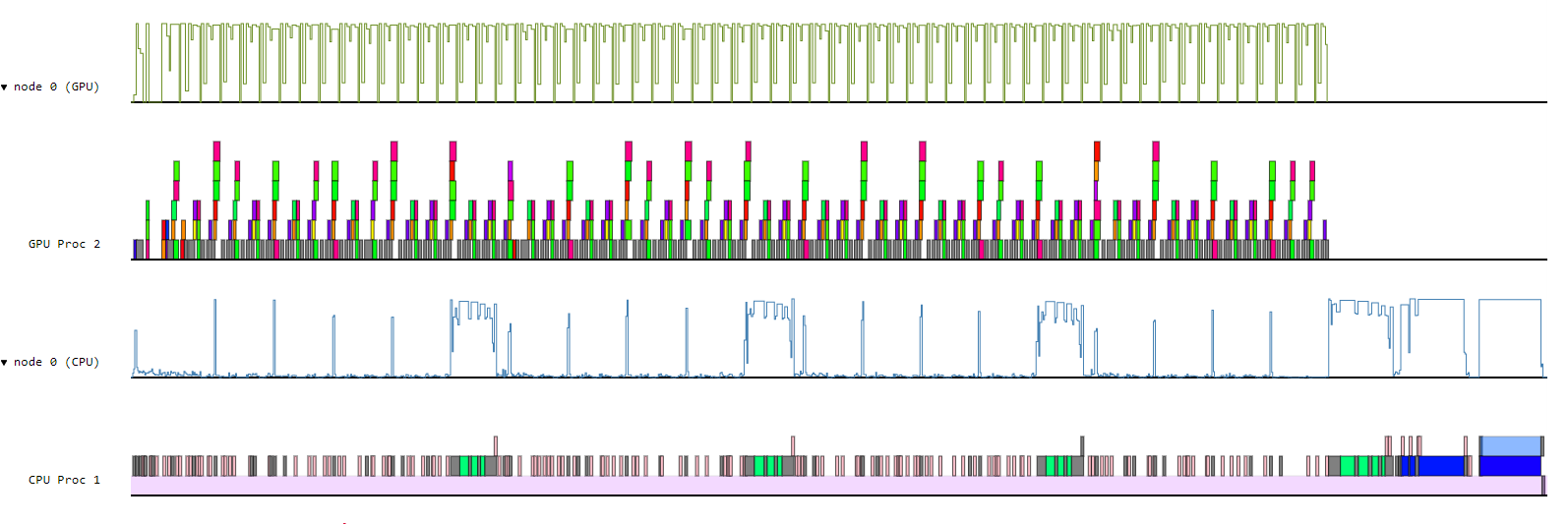}}
      
       \subfloat[]
       {\label{spy_b} \includegraphics[width=1.0\textwidth]{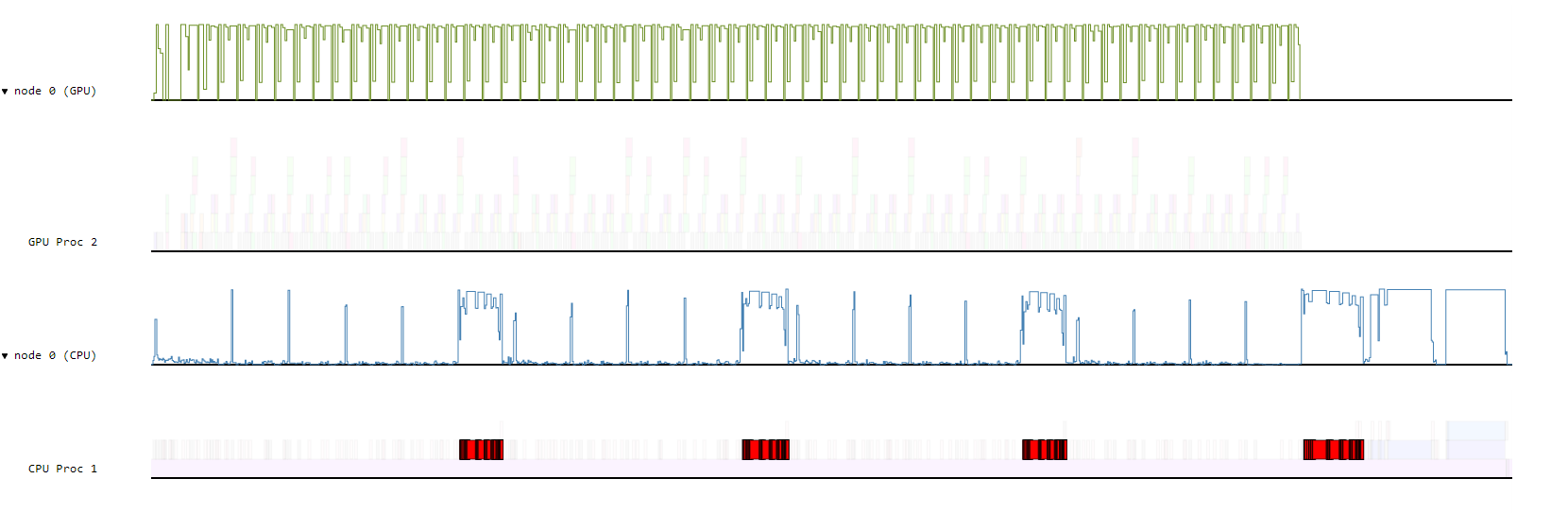}}

       \caption{ \label{fig:cpu-gpu_map} Mapping of the NSID code onto a heterogeneous architecture; the flow solver is mapped to a GPU, and the ID is mapped to a CPU. Figure \protect\subref{spy_a} shows the overall LegionProf profile while Figure \protect\subref{spy_b} highlights stage one of the SPID algorithm.}
      \end{figure*}

 \textit{Portability:} Though the scaling results presented in the previous section were for CPU-only runs, Legion allows the user to port their code to multiple types of processors. The ability to take advantage of this is useful for the paired fluid dynamics-data compression algorithm being discussed here, because there are different characteristics between the two algorithms. The flow solver tasks have low memory requirements, are computation intensive, and can be executed concurrently across each sub-domain of our partition. \heather{These characteristics} are well-suited to GPU architectures. However, data compression algorithms are the exact opposite; memory requirements can be high, especially if large numbers of time steps are being analyzed, and they consist of linear algebra computations that work in serial across the data matrix. Because of this, they are more suited for CPU-type architectures. To address this disparity, a custom mapper within the Legion framework was developed to direct the placement of the fluid solve and the data compression tasks on heterogeneous architectures. Figure~\ref{fig:cpu-gpu_map} shows an example profile for 25 time steps of the NSID solver mapped in this manner.  As expected, the first stage data compression process begins while the flow solver is still running. 
      
When using a homogeneous architecture, e.g., strictly CPUs, any reduction in the runtime of the NSID solver due to Stage 1 of the ID algorithm will be due to the distributed memory nature of the machine. The typical mapping for this scenario is to have one domain partition per processor; latency in the solver occurs due to the exchange of fringe point data across the machine network. If there is enough latency in the system, the Stage 1 tasks can be computed during this time and increase efficiency. However, there is no way to improve the solver performance within a single node. This changes if a user can access and use the entire heterogeneous architecture. By using Legion's mapping capabilities to port all flow solver tasks to a GPU that was previously not accessible to the user, Stage 1 of the SPID algorithm truly becomes parallel with the flow solver; essentially, this stage of the compression is ``free". 

\begin{figure}[!h]
       \center{\includegraphics[width=0.49\textwidth]
       {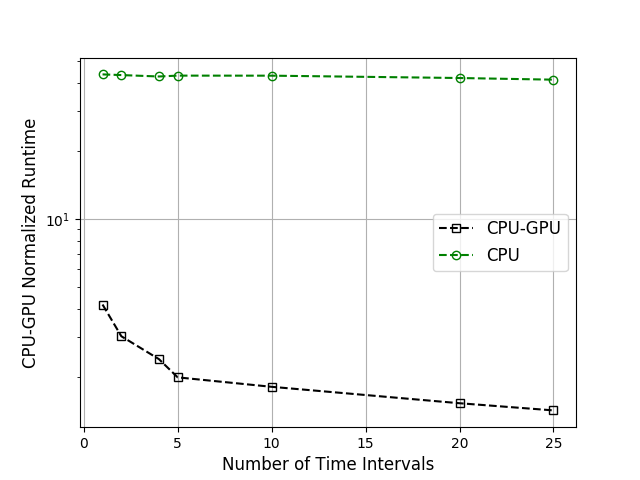}}
       \caption{\label{fig:CPU_GPU_vs_CPU_runtimes} Runtimes normalized against a 1 GPU N-S solve runtime of 5.26 seconds. The problem is solved on a $64^3$ domain for 1 block and 100 time-steps. In the CPU-GPU setup, the GPU is used for the solver and the CPU is used for data compression.}
      \end{figure}

Figure~\ref{fig:CPU_GPU_vs_CPU_runtimes} shows an example of this. For a single CPU, there are no savings within a single node for increasing the number of time step intervals analyzed for Stage 1 of the ID. However, if the flow solver is run on a GPU and the data compression is run on a CPU, there are significant runtime savings as the number of time step intervals is increased. Running the flow solver on a GPU instead of a CPU also decreases the overall runtime by more than a factor of 10. This can be extended to the compression of multiple flow QoIs; since there are traditionally many more CPUs than GPUs on heterogeneous architectures, and each QoI can be independently compressed for each sub-domain, it is possible that many flow QoIs could be compressed in the same amount of solver time. This avenue of research is left to future work.

\section{Conclusions and Future Work}
\label{sec:conclusions}
In this paper, we have presented a novel task-based parallel framework for {\it in situ} compression of data generated from large-scale computational fluid dynamics simulations using the matrix interpolative decomposition. This framework, implemented using the Legion programming system, allows the user to achieve significant compression factors of simulation data with reduced disk memory requirements following simulation and a minimal effect on scaling. We verify this claim on two test problems: an analytical Taylor-Green vortex and a large-scale Navier-Stokes solver implemented in Region, a high-level programming language built on top of Legion.

Because Legion implicitly extracts parallelism from user-defined tasks and task dependencies, the CPU and memory usage due to the compression itself, i.e., the computation of the ID on sub-domains of the full simulation, is negligible compared to the cost of the simulation. 

\alec{The custom mapper} \heather{for the Navier-Stokes ID solver discussed in Section~\ref{NSID}} \alec{in its current form} \heather{is a proof of concept and is not optimized for performance at a very large scale.} A more robust custom mapper is being developed; this mapper will be used to further investigate performance for a much larger number of nodes, as well as study the simultaneous compression of multiple flow QoIs. The behavior and performance of ID within ensemble simulations will also be studied. Future work also includes development of newer ID algorithms as open-source software built in Regent. 

\section{Declaration of Competing Interests}
The authors declare that they have no known competing financial interests or personal relationships that could have appeared to influence the work reported in this paper. 

\section*{Acknowledgements}

This work was funded by the United States Department of Energy's National Nuclear Security Administration under the Predictive Science Academic Alliance Program II at Stanford University, Grant 495 DE-NA-0002373. The work of AD was supported by the AFOSR grant FA9550-20-1-0138.

The computations in this paper were performed on the Yellowstone cluster at the Stanford HPC Center, supported through awards from the National Science Foundation, DOD HPCMP, and Office of Naval Research.
\section*{Appendix}
\setcounter{equation}{0}
\renewcommand\thesection{A}
\renewcommand\thesubsection{\thesection.\Alph{subsection}}
\renewcommand{\theequation}{A.\arabic{equation}}

The following lemma enumerates some of the properties of the ID.
\begin{lemma}
\label{lem:IDproplemma}
(Lemma 3.1 of~\cite{martinssona2011randomized}). For a matrix $\bm{A} \in \mathbb{R}^{m \times n}$ there exists a set $\mathcal{I}\subseteq\{0,\dots,n-1\}$ with $\vert \mathcal{I} \vert = k$, a corresponding column skeleton $\bm{A}(:,\mathcal{I}) \in \mathbb{R}^{m \times k}$, and coefficient matrix $\bm{C} \in \mathbb{R}^{k \times n}$ which satisfies the following properties.
\begin{enumerate}
    \item $\bm{C}(:,\mathcal{I})$ is the $k \times k$ identity matrix.
    \item no entry of $\bm{C}$ has absolute value greater than 1.
    \item $\Vert \bm{C} \Vert_2 \leq \sqrt{k(n-k) + 1}$.
    \item the $k^{th}$ largest singular value of $\bm{C}$, $\sigma_k$, is at least one.
    \item $\bm{A} = \bm{A}(:,\mathcal{I})\bm{C}$ when $k=m$ or $k=n$.
    \item $\Vert \bm{A} - \bm{A}(:,\mathcal{I})\bm{C} \Vert_2 \leq \sqrt{k(n-k) + 1} \sigma_{k+1} $.
    \end{enumerate}
\end{lemma}
The following two theorems provide error estimates for SubID and SPID. Note that Theorem~\ref{thm:subID} is essentially a restatement of Theorem 1 from~\cite{hampton2018practical}.
\begin{theorem}
\label{thm:subID}
(Theorem 1 of~\cite{dunton2020pass}).
Let $\bm{A}$ be a PDE data matrix, $\bm{B} = \bm{A}(\mathcal{J},:)$ the subsampled (coarsened) matrix, $\hat{\bm{B}}$ the rank $k$ ID approximation to $\bm{B}$ as in (\ref{eq:coarseA_ID}), and $\hat{\bm{A}}$ the subsampled ID approximation as in (\ref{eq:coarseregA}). For any $\tau \geq 0$, let 
\begin{equation}
    \epsilon(\tau) := \lambda_{\max}(\bm{A}\bm{A}^T - \tau\bm{B}\bm{B}^T),
    \label{eq:epstau}
\end{equation}
where $\lambda_{\max}$ denotes the largest eigenvalue. Then,
\begin{align}
\label{eqn:id_bound}
\Vert \bm{A} - \hat{\bm{A}}\Vert_2 &\leq \min_{\tau, k \leq \mathrm{rank}(\bm{B})} \rho_k(\tau) ,\\
\rho_k(\tau) &:= (1 + \Vert\bm{C}\Vert_2)\sqrt{\tau\sigma_{k+1}^2 + \epsilon(\tau)} \\
&+ \Vert \bm{B} - \hat{\bm{B}} \Vert_2 \sqrt{\tau +  \epsilon(\tau)\sigma_k^{-2}},
\end{align}
where $\sigma_k$ and $\sigma_{k+1}$ are the $k^{th}$ and $(k+1)^{th}$ largest singular values of $\bm{B}$, respectively.
\label{thm:subIDerror}
\end{theorem}

\begin{theorem}
(Theorem 2 of~\cite{dunton2020pass}).
Let $\bm{A}$ be a PDE data matrix, $\bm{B}$  the subsampled (coarsened) matrix, $\bm{M}$ the interpolation operator as in (\ref{eq:single_ID}) and with associated interpolation error $\bm{E}_I:=\bm{A} - \bm{M}\bm{B}$. Then, the error of the single-pass ID approximation $\hat{\bm{A}}$ in (\ref{eq:single_ID}) is bounded as follows
\begin{align}
\Vert \bm{A} - \hat{\bm{A}}\Vert_2 &\leq \Vert \bm{E}_I \Vert_2 + \Vert \bm{M} \Vert_2 \Vert \bm{B} - \hat{\bm{B}} \Vert_2.
\end{align}
\label{thm:SPIDerror}
\end{theorem}

\bibliographystyle{elsarticle-num}
\bibliography{references}

\begin{thebibliography}{10}
\expandafter\ifx\csname url\endcsname\relax
  \def\url#1{\texttt{#1}}\fi
\expandafter\ifx\csname urlprefix\endcsname\relax\def\urlprefix{URL }\fi
\expandafter\ifx\csname href\endcsname\relax
  \def\href#1#2{#2} \def\path#1{#1}\fi

\bibitem{ang2012report}
J.~Ang, K.~Evans, A.~Geist, M.~Heroux, P.~Hovland, O.~Marques, L.~McInnes,
  E.~Ng, S.~Wild, Report on the workshop on extreme-scale solvers: Transitions
  to future architectures, Office of Advanced Scientific Computing Research, US
  Department of Energy (2012) 8--9.

\bibitem{amarasinghe2009exascale}
S.~Amarasinghe, D.~Campbell, W.~Carlson, A.~Chien, W.~Dally, E.~Elnohazy,
  M.~Hall, R.~Harrison, W.~Harrod, K.~Hill, et~al., Exascale software study:
  Software challenges in extreme scale systems, DARPA IPTO, Air Force Research
  Labs, Tech. Rep (2009) 1--153.

\bibitem{ashby2010opportunities}
S.~Ashby, P.~Beckman, J.~Chen, P.~Colella, B.~Collins, D.~Crawford,
  J.~Dongarra, D.~Kothe, R.~Lusk, P.~Messina, et~al., The opportunities and
  challenges of exascale computing, Summary Report of the Advanced Scientific
  Computing Advisory Committee (ASCAC) Subcommittee (2010) 1--77.

\bibitem{sprague2017turbulent}
M.~A. Sprague, S.~Boldyrev, P.~Fischer, R.~Grout, W.~I. Gustafson~Jr, R.~Moser,
  Turbulent flow simulation at the exascale: Opportunities and challenges
  workshop: August 4-5, 2015, washington, dc, Tech. rep., National Renewable
  Energy Lab.(NREL), Golden, CO (United States) (2017).

\bibitem{asch2018big}
M.~Asch, T.~Moore, R.~Badia, M.~Beck, P.~Beckman, T.~Bidot, F.~Bodin,
  F.~Cappello, A.~Choudhary, B.~de~Supinski, et~al., Big data and extreme-scale
  computing: Pathways to convergence-toward a shaping strategy for a future
  software and data ecosystem for scientific inquiry, The International Journal
  of High Performance Computing Applications 32~(4) (2018) 435--479.

\bibitem{gerber2018crosscut}
R.~Gerber, J.~Hack, K.~Riley, K.~Antypas, R.~Coffey, E.~Dart, T.~Straatsma,
  J.~Wells, D.~Bard, S.~Dosanjh, et~al., Crosscut report: Exascale requirements
  reviews, march 9--10, 2017--tysons corner, virginia. an office of science
  review sponsored by: Advanced scientific computing research, basic energy
  sciences, biological and environmental research, fusion energy sciences, high
  energy physics, nuclear physics, Tech. rep., Oak Ridge National Lab.(ORNL),
  Oak Ridge, TN (United States); Argonne~… (2018).

\bibitem{kunkel2014exascale}
J.~M. Kunkel, M.~Kuhn, T.~Ludwig, Exascale storage systems: an analytical study
  of expenses, Supercomputing frontiers and innovations 1~(1) (2014) 116--134.

\bibitem{rasquin2014scalable}
M.~Rasquin, C.~Smith, K.~Chitale, E.~S. Seol, B.~Matthews, J.~Martin, O.~Sahni,
  R.~Loy, M.~S. Shephard, K.~E. Jansen, Scalable fully implicit finite element
  flow solver with application to high-fidelity flow control simulations on a
  realistic wing design, Computing in Science and Engineering 16~(6) (2014)
  13--21.

\bibitem{li2018data}
S.~Li, N.~Marsaglia, C.~Garth, J.~Woodring, J.~Clyne, H.~Childs, Data reduction
  techniques for simulation, visualization and data analysis, in: Computer
  Graphics Forum, Vol.~37, Wiley Online Library, 2018, pp. 422--447.

\bibitem{gong2012multi}
Z.~Gong, S.~Lakshminarasimhan, J.~Jenkins, H.~Kolla, S.~Ethier, J.~Chen,
  R.~Ross, S.~Klasky, N.~F. Samatova, Multi-level layout optimization for
  efficient spatio-temporal queries on isabela-compressed data, in: 2012 IEEE
  26th International Parallel and Distributed Processing Symposium, IEEE, 2012,
  pp. 873--884.

\bibitem{fowler1994lossless}
J.~E. Fowler, R.~Yagel, Lossless compression of volume data, in: Proceedings of
  the 1994 symposium on Volume visualization, 1994, pp. 43--50.

\bibitem{di2016fast}
S.~Di, F.~Cappello, Fast error-bounded lossy hpc data compression with sz, in:
  2016 ieee international parallel and distributed processing symposium
  (ipdps), IEEE, 2016, pp. 730--739.

\bibitem{tao2017significantly}
D.~Tao, S.~Di, Z.~Chen, F.~Cappello, Significantly improving lossy compression
  for scientific data sets based on multidimensional prediction and
  error-controlled quantization, in: 2017 IEEE International Parallel and
  Distributed Processing Symposium (IPDPS), IEEE, 2017, pp. 1129--1139.

\bibitem{liang2018error}
X.~Liang, S.~Di, D.~Tao, S.~Li, S.~Li, H.~Guo, Z.~Chen, F.~Cappello,
  Error-controlled lossy compression optimized for high compression ratios of
  scientific datasets, in: 2018 IEEE International Conference on Big Data (Big
  Data), IEEE, 2018, pp. 438--447.

\bibitem{lindstrom2006fast}
P.~Lindstrom, M.~Isenburg, Fast and efficient compression of floating-point
  data, IEEE transactions on visualization and computer graphics 12~(5) (2006)
  1245--1250.

\bibitem{ibarria2003out}
L.~Ibarria, P.~Lindstrom, J.~Rossignac, A.~Szymczak, Out-of-core compression
  and decompression of large n-dimensional scalar fields, in: Computer Graphics
  Forum, Vol.~22, Wiley Online Library, 2003, pp. 343--348.

\bibitem{lakshminarasimhan2011compressing}
S.~Lakshminarasimhan, N.~Shah, S.~Ethier, S.~Klasky, R.~Latham, R.~Ross, N.~F.
  Samatova, Compressing the incompressible with isabela: In-situ reduction of
  spatio-temporal data, in: European Conference on Parallel Processing,
  Springer, 2011, pp. 366--379.

\bibitem{lehmann2014situ}
H.~Lehmann, B.~Jung, In-situ multi-resolution and temporal data compression for
  visual exploration of large-scale scientific simulations, in: 2014 ieee 4th
  symposium on large data analysis and visualization (ldav), IEEE, 2014, pp.
  51--58.

\bibitem{otero2018lossy}
E.~Otero, R.~Vinuesa, O.~Marin, E.~Laure, P.~Schlatter, Lossy data compression
  effects on wall-bounded turbulence: bounds on data reduction, Flow,
  Turbulence and Combustion 101~(2) (2018) 365--387.

\bibitem{marin2016large}
O.~Marin, M.~Schanen, P.~Fischer, Large-scale lossy data compression based on
  an a priori error estimator in a spectral element code, Tech. rep.,
  ANL/MCS-P6024-0616 (2016).

\bibitem{yeo1995volume}
B.-L. Yeo, B.~Liu, Volume rendering of dct-based compressed 3d scalar data,
  IEEE Transactions on Visualization and Computer Graphics 1~(1) (1995) 29--43.

\bibitem{cohen1992biorthogonal}
A.~Cohen, I.~Daubechies, J.-C. Feauveau, Biorthogonal bases of compactly
  supported wavelets, Communications on pure and applied mathematics 45~(5)
  (1992) 485--560.

\bibitem{farge1992wavelet}
M.~Farge, Wavelet transforms and their applications to turbulence, Annual
  review of fluid mechanics 24~(1) (1992) 395--458.

\bibitem{strang1996wavelets}
G.~Strang, T.~Nguyen, Wavelets and filter banks, SIAM, 1996.

\bibitem{lindstrom2014fixed}
P.~Lindstrom, Fixed-rate compressed floating-point arrays, IEEE transactions on
  visualization and computer graphics 20~(12) (2014) 2674--2683.

\bibitem{loeve1977elementary}
M.~Loeve, Elementary probability theory, in: Probability Theory I, Springer,
  1977, pp. 1--52.

\bibitem{therrien1992discrete}
C.~W. Therrien, Discrete random signals and statistical signal processing,
  Prentice Hall PTR, 1992.

\bibitem{hitchcock1927expression}
F.~L. Hitchcock, The expression of a tensor or a polyadic as a sum of products,
  Journal of Mathematics and Physics 6~(1-4) (1927) 164--189.

\bibitem{tucker1966some}
L.~R. Tucker, Some mathematical notes on three-mode factor analysis,
  Psychometrika 31~(3) (1966) 279--311.

\bibitem{kroonenberg1980principal}
P.~M. Kroonenberg, J.~De~Leeuw, Principal component analysis of three-mode data
  by means of alternating least squares algorithms, Psychometrika 45~(1) (1980)
  69--97.

\bibitem{de2000best}
L.~De~Lathauwer, B.~De~Moor, J.~Vandewalle, On the best rank-1 and rank-(r 1, r
  2,..., rn) approximation of higher-order tensors, SIAM journal on Matrix
  Analysis and Applications 21~(4) (2000) 1324--1342.

\bibitem{vannieuwenhoven2012new}
N.~Vannieuwenhoven, R.~Vandebril, K.~Meerbergen, A new truncation strategy for
  the higher-order singular value decomposition, SIAM Journal on Scientific
  Computing 34~(2) (2012) A1027--A1052.

\bibitem{austin2016parallel}
W.~Austin, G.~Ballard, T.~G. Kolda, Parallel tensor compression for large-scale
  scientific data, in: 2016 IEEE international parallel and distributed
  processing symposium (IPDPS), IEEE, 2016, pp. 912--922.

\bibitem{glaws2020deep}
A.~Glaws, R.~King, M.~Sprague, Deep learning for in situ data compression of
  large turbulent flow simulations, Physical Review Fluids 5~(11) (2020)
  114602.

\bibitem{azaiez2019low}
M.~Aza{\"\i}ez, L.~Lestandi, T.~C. Rebollo, Low rank approximation of
  multidimensional data, in: High-Performance Computing of Big Data for
  Turbulence and Combustion, Springer, 2019, pp. 187--250.

\bibitem{brand2006fast}
M.~Brand, Fast low-rank modifications of the thin singular value decomposition,
  Linear algebra and its applications 415~(1) (2006) 20--30.

\bibitem{zimmermann2018geometric}
R.~Zimmermann, B.~Peherstorfer, K.~Willcox, Geometric subspace updates with
  applications to online adaptive nonlinear model reduction, SIAM Journal on
  Matrix Analysis and Applications 39~(1) (2018) 234--261.

\bibitem{tropp2019streaming}
J.~A. Tropp, A.~Yurtsever, M.~Udell, V.~Cevher, Streaming low-rank matrix
  approximation with an application to scientific simulation, SIAM Journal on
  Scientific Computing 41~(4) (2019) A2430--A2463.

\bibitem{cheng2005compression}
H.~Cheng, Z.~Gimbutas, P.-G. Martinsson, V.~Rokhlin, On the compression of low
  rank matrices, SIAM Journal on Scientific Computing 26~(4) (2005) 1389--1404.

\bibitem{halko2011finding}
N.~Halko, P.-G. Martinsson, J.~A. Tropp, Finding structure with randomness:
  Probabilistic algorithms for constructing approximate matrix decompositions,
  SIAM review 53~(2) (2011) 217--288.

\bibitem{dunton2020pass}
A.~M. Dunton, L.~Jofre, G.~Iaccarino, A.~Doostan, Pass-efficient methods for
  compression of high-dimensional turbulent flow data, Journal of Computational
  Physics 423 (2020) 109704.

\bibitem{Legion_paper}
M.~Bauer, S.~Treichler, E.~Slaughter, A.~Aiken, Legion: Expressing locality and
  independence with logical regions, in: Supercomputing Conference (SC12),
  2012.

\bibitem{yu2017single}
W.~Yu, Y.~Gu, J.~Li, S.~Liu, Y.~Li, Single-pass pca of large high-dimensional
  data, arXiv preprint arXiv:1704.07669 (2017).

\bibitem{martinsson2019randomized}
P.-G. Martinsson, Randomized methods for matrix computations, The Mathematics
  of Data 25  187--231.

\bibitem{golub2012matrix}
G.~H. Golub, C.~F. Van~Loan, Matrix computations, Vol.~3, JHU press, 2012.

\bibitem{gu1996efficient}
M.~Gu, S.~C. Eisenstat, Efficient algorithms for computing a strong
  rank-revealing qr factorization, SIAM Journal on Scientific Computing 17~(4)
  (1996) 848--869.

\bibitem{mahoney2009cur}
M.~W. Mahoney, P.~Drineas, Cur matrix decompositions for improved data
  analysis, Proceedings of the National Academy of Sciences 106~(3) (2009)
  697--702.

\bibitem{elhamifar2009sparse}
E.~Elhamifar, R.~Vidal, Sparse subspace clustering, in: 2009 IEEE Conference on
  Computer Vision and Pattern Recognition, IEEE, 2009, pp. 2790--2797.

\bibitem{dyer2015self}
E.~L. Dyer, T.~A. Goldstein, R.~Patel, K.~P. Kording, R.~G. Baraniuk,
  Self-expressive decompositions for matrix approximation and clustering, arXiv
  preprint arXiv:1505.00824 (2015).

\bibitem{perry2019allocation}
D.~J. Perry, R.~M. Kirby, A.~Narayan, R.~T. Whitaker, Allocation strategies for
  high fidelity models in the multifidelity regime, SIAM/ASA Journal on
  Uncertainty Quantification 7~(1) (2019) 203--231.

\bibitem{liberty2007randomized}
E.~Liberty, F.~Woolfe, P.-G. Martinsson, V.~Rokhlin, M.~Tygert, Randomized
  algorithms for the low-rank approximation of matrices, Proceedings of the
  National Academy of Sciences 104~(51) (2007) 20167--20172.

\bibitem{woodruff2014sketching}
D.~P. Woodruff, Sketching as a tool for numerical linear algebra, arXiv
  preprint arXiv:1411.4357 (2014).

\bibitem{parallel_text}
T.~Rauber, G.~R{\"u}nger, Parallel Programming for Multicore and Cluster
  Systems, Springer, 2010.

\bibitem{pi2013scalable}
Y.~Pi, H.~Peng, S.~Zhou, Z.~Zhang, A scalable approach to column-based low-rank
  matrix approximation, in: Proceedings of the Twenty-Third international joint
  conference on Artificial Intelligence, 2013, pp. 1600--1606.

\bibitem{Runtime_paper}
S.~Treichler, M.~Bauer, A.~Aiken, Realm: An event-based low-level runtime for
  distributed memory architectures, in: Parallel Architectures and Compilation
  Techniques (PACT 2014), 2014.

\bibitem{PAW-ATM}
Parallel applications workshop, alternatives to mpi+x, in: SC19: The
  International Conference for High Performance Computing, Networking, Storage,
  and Analysis, IEEE, 2019.

\bibitem{spy_ex}
\href{https://legion.stanford.edu/debugging/}{Legion programming system:
  Debugging}.
\newline\urlprefix\url{https://legion.stanford.edu/debugging/}

\bibitem{Regent_paper}
E.~Slaughter, W.~Lee, S.~Treichler, M.~Bauer, A.~Aiken, Regent: A
  high-productivity programming language for hpc with logical regions, in:
  Supercomputing Conference (SC15), 2015.

\bibitem{cart_p_paper_2}
A.~Wissink, S.~K. andT. Pulliam, J.~Sitaramen, V.~Sankaren, Cartesian adaptive
  mesh refinement for rotorcraft wake resolution, in: The 28th AIAA Applied
  Aerodynamics Conference, 2010.

\bibitem{cart_p_paper}
J.~Leffell, J.~Sitaramen, V.~Lakshminarayan, A.~Wissink, Towards efficient
  parallel-in-time simulation of periodic flow, in: The 54th AIAA Aerospace
  Sciences Meeting, 2016.

\bibitem{rk3}
C.~Kennedy, M.~Carpenter, R.~Lewis, Low-storage, explicit runge–kutta schemes
  for the compressible navier–stokes equations, Applied Numerical Mathematics
  35 (2000) 177--219.

\bibitem{martinssona2011randomized}
P.-G. Martinssona, V.~Rokhlin, M.~Tygertc, A randomized algorithm for the
  decomposition of matrices, Appl. Comput. Harmon. Anal 30 (2011) 47--68.

\bibitem{hampton2018practical}
J.~Hampton, H.~R. Fairbanks, A.~Narayan, A.~Doostan, Practical error bounds for
  a non-intrusive bi-fidelity approach to parametric/stochastic model
  reduction, Journal of Computational Physics 368 (2018) 315--332.

\end{thebibliography}

\end{document}